\documentclass{article}
\usepackage{PRIMEarxiv}
\usepackage{booktabs}
\usepackage{graphicx}

\usepackage{makecell}
\usepackage{tabularx}
\newcolumntype{C}[1]{>{\centering\arraybackslash}p{#1}}

\usepackage{multirow}
\usepackage[english]{babel}
\usepackage{amsmath} 
\usepackage{amssymb}
\usepackage[table]{xcolor}
\usepackage{siunitx}
\usepackage{algorithm}
\usepackage{algpseudocode}
\usepackage{caption}
\usepackage{amsfonts}
\usepackage{listings}
\usepackage{parskip}
\usepackage{float}
\usepackage{wrapfig}
\usepackage{needspace}
\usepackage{geometry}
\usepackage[authoryear,round]{natbib}
\usepackage{hyperref}
\usepackage{mathtools}
\usepackage{adjustbox}
\usepackage{siunitx}
\usepackage{subcaption}
\usepackage[utf8]{inputenc} 
\usepackage[T1]{fontenc}    
\usepackage{url}            
\usepackage{nicefrac}       
\usepackage{microtype}      
\usepackage{fancyhdr}       
\graphicspath{{media/}}     
\usepackage{comment}
\usepackage{cleveref}
\usepackage{textcomp}
\usepackage{tcolorbox}
\usepackage{tikz}
\usepackage{tikz-3dplot}
\usepackage{pifont}
\usepackage{csquotes}
\usepackage{scalerel}
\usepackage{stackengine,wasysym}
\usepackage{xcolor}
\usepackage{colortbl}
\usepackage{booktabs}
\usepackage{fontawesome5}
\graphicspath{{media/}{visuals/}}     % organize your images and other figures under media/ folder

\usetikzlibrary{arrows.meta, calc, backgrounds, positioning}
\usetikzlibrary{shapes,arrows,calc,positioning,shadings}

\definecolor{codegreen}{rgb}{0,0.6,0}
\definecolor{codegray}{rgb}{0.5,0.5,0.5}
\definecolor{codepurple}{rgb}{0.58,0,0.82}
\definecolor{backcolour}{rgb}{0.95,0.95,0.92}
\definecolor{srcblue}{RGB}{18,62,138}
\definecolor{estorange}{RGB}{201,110,17}
\newcommand{\src}[1]{\textcolor{srcblue}{#1}}
\newcommand{\est}[1]{\textcolor{estorange}{#1}}

\definecolor{ForestGreen}{RGB}{34,139,34}
\definecolor{MidnightBlue}{RGB}{25, 25, 112}

\usetikzlibrary{shapes,arrows,positioning}

\hypersetup{
    colorlinks=true,
    linktocpage=true,
    urlcolor=MidnightBlue,
    linkcolor=red,
    citecolor=ForestGreen
}

\definecolor{bg}{gray}{0.95}

 \newcommand\blfootnote[1]{%
  \begingroup
  \renewcommand\thefootnote{}\footnote{#1}%
  \addtocounter{footnote}{-1}%
  \endgroup
}

\def\sym#1{\ifmmode^{#1}\else\(^{#1}\)\fi}

\definecolor{headerblue}{RGB}{234,240,250}
\newcommand{\ph}[1]{\textcolor{red}{#1}}

\renewcommand{\topfraction}{0.80}
\renewcommand{\bottomfraction}{0.30}
\renewcommand{\textfraction}{0.20}
\renewcommand{\floatpagefraction}{0.65}
\title{VoiceMem: Streaming Dual-Brain Memory \\for Real-Time Interaction}
\vspace{-16mm}
\small
\author{
Zhifei Xie$^{1,5*}$ \quad
Jiaqi Lang$^{2*}$ \quad
Ze An$^{2}$ \quad
Yifan Zhao$^{2}$ \quad
Dongchao Yang$^{4}$ \quad
Kai Li$^{3}$ \\
Ziyang Ma$^{1,5}$ \quad 
Mingbao Lin$^{2\dagger}$ \quad
Chunyan Miao$^{1\dagger}$ \quad
Shuicheng Yan$^{2\dagger}$ \\[0.35em]
{\small
$^{1}$Nanyang Technological University \quad
$^{2}$National University of Singapore \quad
$^{3}$Tsinghua University
}\\[-0.05em]
{\small
$^{4}$The Chinese University of Hong Kong \quad
$^{5}$Open Interaction Lab
}\\[-0.1em]
{\small \faEnvelope\ \texttt{Zhifei001@e.ntu.edu.sg}}
}

\begin{document}

\maketitle
\vspace{-10.5mm}

\definecolor{darkpink}{RGB}{225, 95, 150}

\begin{center}
\href{https://xzf-thu.github.io/VoiceMem/}
{\textcolor{darkpink}{\textbf{\texttt{https://xzf-thu.github.io/VoiceMem/}}}}
\end{center}

\vspace{-2mm}
\begin{figure}[!h]
    \centering
    \includegraphics[width=0.98\linewidth]{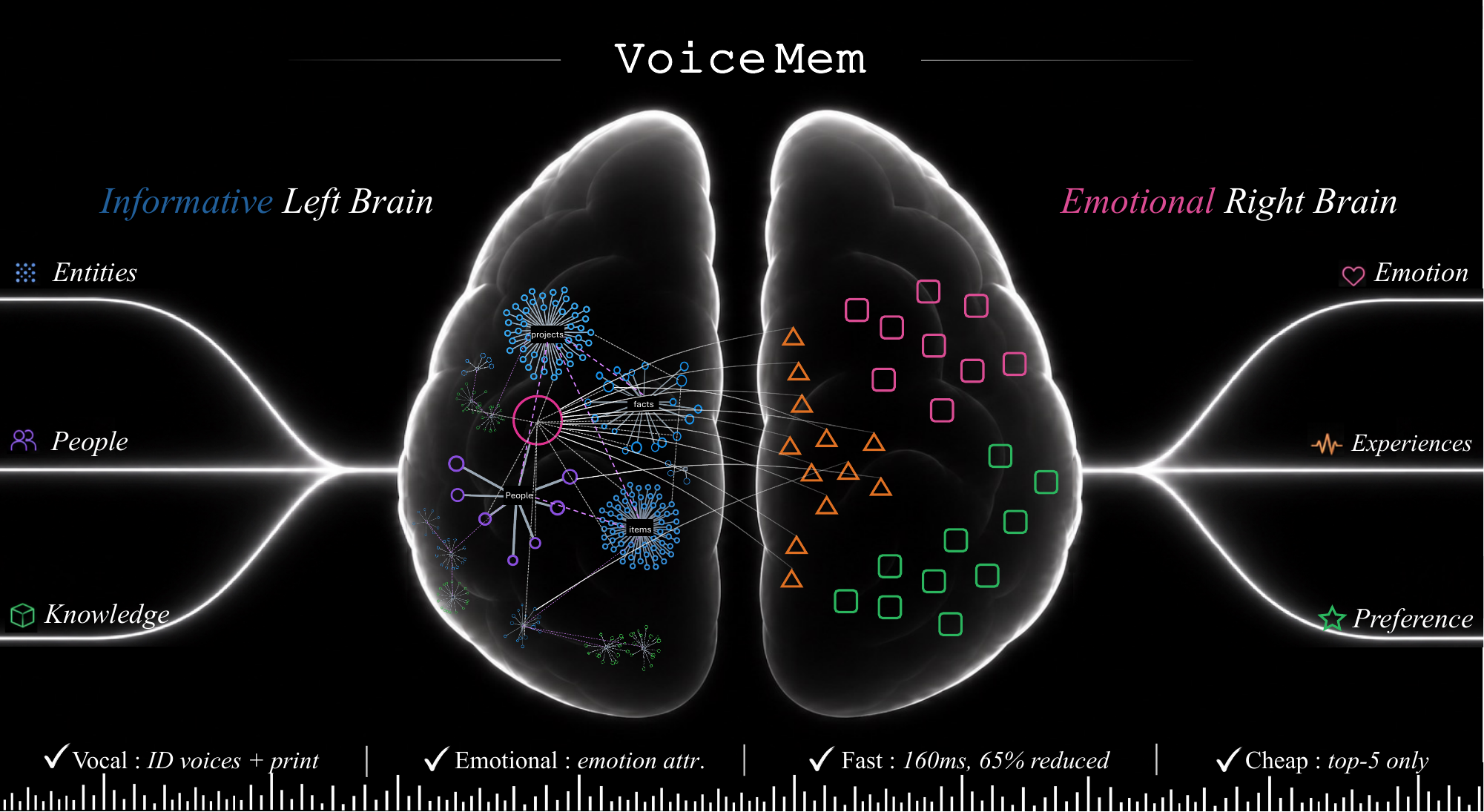}
    \vspace{-1mm}
    \caption{\textsc{VoiceMem}: a memory system purpose-built for real-time conversational systems. The ``left brain'' provides state-of-the-art memory capabilities, while the ``right brain'' captures emotion and persona. \textbf{Together, they enable rich, personalized memory with virtually no added latency.}}
    \label{fig:1-teaser}
\end{figure}
\vspace{4mm}

\begin{abstract}

% 交互系统（eg, duplex Speech Languagde Models）仍然缺乏一个流式，精准且有同理心的记忆系统作为 its soul。我们提出一个非常简单的记忆架构，VoiceMem, 包含并行的informational左脑和emotional右脑，以及流式的存取机制。进一步地，我们补全了model training pipeline for 让SLMs可以接受audio-memory输入，validation for 长期对话的对话系统以及可持续进化的解耦部署infrastrater。 实验和工程实践证明了我们达到了基本的目标： （1） 加粗accuracy：on left brain，voicemem在top-5 setting下超过经典的Mem0等system on top-200， 涨幅超过22%。 （2）emotional & personal： 结合右脑的长-短期情绪归因机制和双节点设计，在5个benchmark上依然维持sota，综合成绩上升16% compared to previous sota。（3）Real-time and Cheap: VoiceMem在常规的vad setting下即可完成所有查询，不增加对话延迟，并在top-5setting下保证精准度和低成本。
Conversational systems, such as duplex speech language models (SLMs), still lack a \underline{\textbf{\textit{streaming, accurate, and empathetic}}} memory system as their soul. We introduce \textbf{\textsc{VoiceMem}}, a simple memory architecture with a parallel informational \textbf{\emph{left brain}}, an emotional \textbf{\emph{right brain}}, and \textbf{streaming memory I/O mechanisms}. We further build a complete pipeline for memory-aware SLM training, long-horizon evaluation, and decoupled deployment with interchangeable memory backends. Experiments and real-world deployment show three advantages: \textbf{i) Accuracy:} under top-$5$ retrieval, the left brain outperforms classical systems such as \textsc{Mem0} at top-$200$ by nearly $30$ points; \textbf{ii) Emotional \& Personal:} the right brain, with short- and long-horizon affective attribution and dual-node persona modeling, achieves state-of-the-art performance across three persona benchmarks and improves the aggregate score by $1.89$ points over the previous best system; and \textbf{iii) Real-Time \& Cheap:} \textsc{VoiceMem} completes retrieval in \underline{\textbf{134\,ms}}, well within standard VAD latency, adding no extra conversational delay while maintaining high accuracy and low cost. These results show that \textsc{VoiceMem} provides a practical memory foundation for real-time, personalized, and emotionally aware speech interaction.

\end{abstract}

\blfootnote{}

\vspace{-10mm}
\section{Introduction}
\label{sec:introduction}

\noindent
\hspace*{10mm}
\begin{minipage}{0.9\textwidth}
\textbf{\textit{"Through memory, the soul reveals traces of its former existence."}}
\end{minipage}

\vspace{-4mm} % 调上下位置
\noindent
\hspace*{120mm} % 调左右位置
\textbf{\textit{— Plato}}
% An assistant that answers well but forgets everything between sessions stays a
% tool. Memory is what lets it become a partner~\citep{park2023generativeagents,
% packer2023memgpt, zhong2023memorybank}. Two lines of work have each solved half
% of this problem. Mem0, Zep, and A-MEM build durable structured stores over long
% dialogue histories~\citep{chhikara2025mem0, rasmussen2025zep, xu2025amem,
% apexmem2026}. Moshi, LSLM, and SyncLLM let a model listen and speak at the same
% time, at human timing~\citep{defossez2024moshi, ma2024lslm, veluri2024syncllm,
% salmonn2024omni, duplexsla2026}. Nobody has put the two together. A real-time
% voice agent still has no memory layer it can afford to query, and nowhere to
% keep what it learns about the person it is talking to.

\vspace{-1mm}
Memory is what turns a conversational system (e.g. voice agent) from an intelligent tool into a human-centered partner. Although many powerful memory systems have appeared recently~\citep{mem0, a-mem, evermemos}, alongside interaction models that are increasingly intelligent and natural (e.g. SLMs~\citep{xie2024mini, xie2026audio, llama-omni,qwen3.5-omni} and duplex models~\citep{lslm, seed-realtime, gpt-live}), the two have not been merged into one complete solution.  Building an empathetic and intelligent interaction system on top of memory remains an open challenge.

% As we see it, the field faces three core challenges. \textbf{(C1) Architecture.} A unified architecture has to carry human-centered information and "EQ" as parallel systems, not simply file emotional information into the factual store. \textbf{(C2) Density under latency.} Memory and affective information have to be extracted at high density under ultra-low retrieval latency, so that a streaming dialogue system stays natural. \textbf{(C3) Evolvability.} The engineering has to keep adapting to the latest mainstream memory backends and the latest interaction models. Native multimodal capability should be folded in as well, as an extra source of support.

From our view, three obstacles stand in the way. \underline{\textbf{(O1) Unified architecture for both informational }}
\underline{\textbf{and emotional intelligence.}} For real-time conversational systems, a grasp of emotion and persona matters as much as information and intelligence, and the mechanisms behind emotion are far more complex. How to model both in one architecture over the long run is still unexplored. \underline{\textbf{(O2) High}} \underline{\textbf{information density under zero latency.}} Existing memory systems conflict with conversational systems in two ways: \textbf{\textit{(i) a \textbf{2--3\,s} retrieval breaks the \textbf{500\,ms} budget of a real-time conversation, and (ii) the \textbf{top-100} output common in text agent memory is more than a speech model can take.}} Instead, a memory system has to add almost no latency, run streaming, and still hold its accuracy at \textbf{top-5}. \underline{\textbf{(O3) Infrastructure and evolvability.}} Memory methods and speech dialogue models are both moving fast. Dialogue models still cannot take joint audio-memory input, and a memory system has to stay simple-and-effective to remain useful and keep updating in a fast-moving field.

\underline{\textbf{We present \textbf{\textsc{VoiceMem}}, a \textit{simple-and-effective} streaming dual-brain memory framework for}} \underline{\textbf{real-time spoken interaction}}, As shown in Figure~\ref{fig:1-teaser}. For \textbf{O1} and \textbf{O2}, \textsc{VoiceMem} has a left brain for information and a right brain for emotion and persona. \textit{\textbf{(i) In the left brain}}, a simple two-level \emph{schema--entity} architecture manages and routes memory items, which raises the information density of the ranking stage for \textbf{top-5} performance. The setting of schemas is critical here, so we add an \emph{emergence} mechanism, driven jointly by left-brain state, semantics, and query frequency, to balance the number of memory schemas against precision over the long run. \textbf{\textit{(ii) In the right brain}}, we define two kinds of nodes to model complex human emotion and persona: \emph{independent nodes} for emotional features of the person, and \emph{cross-entity nodes} for emotional features tied to entities in the left brain, both maintained by a short- and long-term emotion attribution mechanism. \textit{\textbf{(iii) To further cut latency}}, we propose a four-stage streaming memory query that breaks every step apart and runs in real time, \textit{finishing the query within the time a standard VAD takes}.

For \textbf{O3}, we reassemble \textsc{VoiceMem} into a fully decoupled framework, with two new modules added. \textit{\textbf{(i)Model adaption:}} To avoid forgetting, we train at small scale and propose \textit{SLM-verified blackbox OPD}, which trains the model through a loop of memory-world construction, online correction by closed-source model, and post-verification. The data produced along the loop gives \textsc{ChatMem-400k} for training and \textsc{ChatMem-Bench} for evaluation, a benchmark covering four dimensions \underline{\textbf{\textit{\{Information, Persona, Affective Attribution, Paralinguist\&Environment\}}}}, across \underline{\textbf{14}} fine-grained categories. \textit{\textbf{(ii) Standalone memory engine:}} Memory methods are moving fast, so we spent considerable extra effort transforming \textsc{VoiceMem} into a \textit{graph-on-graph} framework. The algorithms of \textsc{VoiceMem} only manage, route, and stream at the upper layer; the lower layer is fully independent. We use Mem0 as the lower layer for its current state-of-the-art performance. Our extensive experiments across a wide range of benchmarks spanning \textbf{information memory} (\textbf{+46.1\%} over Mem0 / \textbf{+16.0\%} over the previous SOTA), \textbf{persona memory} (\textbf{+16.8\%} / \textbf{+5.9\%}), and \textbf{long-horizon audio memory} (\textbf{+41.3\%} / \textbf{+27.4\%}) demonstrate that \textsc{VoiceMem} consistently advances memory quality across text and speech, while supporting low-latency streaming retrieval and transferring robustly across different memory backends.

\vspace{-2mm}
\section{Preliminary}

\vspace{-1mm}
\textbf{Embedding-Based Memory Retrieval.} Retrieval-Augmented Generation (RAG) retrieves external
records according to semantic similarity
\citep{rag}. General memory engines such as Mem0~\citep{mem0} and Zep~\citep{zep} further
support memory writing and 
updating. Let $\mathcal{M}_t$ be the memory store at step $t$ and
$q_t$ the current query. Standard retrieval selects
\vspace{-4mm}

\begin{equation}
\mathcal{R}^{\mathrm{sem}}_t
=
\operatorname{TopK}_{m_i\in\mathcal{M}_t}
\operatorname{sim}
\left(f_\theta(q_t),f_\theta(m_i)\right),
\tag{1}
\end{equation}

\vspace{-2mm}
and generates
\begin{equation}
o_t=\operatorname{LLM}
\left(q_t,\mathcal{R}^{\mathrm{sem}}_t\right),
\qquad
\mathcal{M}_{t+1}
=
\operatorname{Update}(\mathcal{M}_t,u_t,o_t).
\tag{2}
\end{equation}

\vspace{-3mm}
Agentic memory systems (e.g., A-Mem~\citep{a-mem} and MemoryOS) and personal memory systems (e.g., MemoryBank~\citep{memorybank} and EverMemOS~\citep{evermemos}) add additional storage structures on top of this base.

\textbf{Emotion-Aware Retrieval.} Emotion-aware systems additionally consider affective
compatibility. Emotional RAG~\citep{emotionalrag} incorporates emotional relevance
into retrieval, while KEEM~\citep{keem}
uses emotional context to guide memory construction and
updating. Let $\mathbf{a}_t$ and $\mathbf{a}_i$ denote the affective
representations of the current query and memory $m_i$.
Retrieval becomes
\begin{equation}
\mathcal{R}^{\mathrm{aff}}_t
=
\operatorname{TopK}_{m_i\in\mathcal{M}_{\star t}}
\left[
\lambda\,\operatorname{sim}\big(
f_{\star\theta}(q_t),\,f_{\star\theta}(m_i)
\big)
+
(1-\lambda)\kappa(\mathbf{a}_t,\mathbf{a}_i)
\right].
\tag{3}
\end{equation}
Despite incorporating affect into retrieval, these methods
remain information-centric, with limited capacity for
emotional association and long-term affective accumulation.

% Emotion-aware systems additionally consider affective
% compatibility. Emotional RAG jointly uses semantic and
% emotional relevance, while KEEM and related systems
% incorporate emotional context into memory construction and
% updating
% \citep{huang2024emotionalrag,kang2025keem,
% lu2025dynamic}.

% Let $\mathbf{a}_t$ and $\mathbf{a}_i$ denote the affective
% representations of the current query and memory $m_i$.
% Retrieval becomes
% %
% \begin{equation}
% \mathcal{R}^{\mathrm{aff}}_t
% =
% \operatorname{TopK}_{m_i\in\mathcal{M}_t}
% \left[
% \lambda\,\operatorname{sim}
% \left(f_\theta(q_t),f_\theta(m_i)\right)
% +
% (1-\lambda)\kappa(\mathbf{a}_t,\mathbf{a}_i)
% \right].
% \tag{3}
% \end{equation}
% %
% This retrieves memories that are both semantically relevant
% and emotionally compatible. However, affect is usually
% treated as metadata or an additional retrieval signal rather
% than an independently evolving cognitive structure.

\textbf{Streaming Dual-Brain Memory} % 信息和感情智能对应着不同的维护方式，一个是为了大量信息的存储，一个强调沉淀和归因。因此，我们extent前作为并行双脑。
Informational and emotional intelligence call for different maintenance: one stores information at scale, the other accumulates and attributes it. We therefore extend prior work into two parallel brains. 
The left brain $\mathcal{G}^{L}_t$ organizes factual cells
into evolving fact clusters, while the right brain
$\mathcal{G}^{R}_t$ maintains affective-attribution cells.
They are connected by cross-brain associations
$\mathcal{E}^{LR}_t$:
\begin{equation}
\mathcal{B}_t=
\left(
\mathcal{G}^{L}_t,
\mathcal{G}^{R}_t,
\mathcal{E}^{LR}_t
\right).
\tag{4}
\end{equation}
For every incoming utterance $u_t$, both brains activate
new cells and update incrementally:
\begin{equation}
\begin{aligned}
\mathcal{G}^{L}_{t+1}
&=
\operatorname{Update}_{L}
\left(
\mathcal{G}^{L}_t,
\operatorname{Activate}_{L}(u_t)
\right),\\
\mathcal{G}^{R}_{t+1}
&=
\operatorname{Update}_{R}
\left(
\mathcal{G}^{R}_t,
\operatorname{Activate}_{R}(u_t)
\right).
\end{aligned}
\tag{5}
\end{equation}
The response is produced through joint retrieval:
\vspace{-5mm}

\begin{equation}
\mathcal{R}^{\mathrm{DB}}_t
=
\operatorname{JointRetrieve}
\left(
q_t\mid
\mathcal{G}^{L}_{t+1},
\mathcal{G}^{R}_{t+1},
\mathcal{E}^{LR}_{t+1}
\right),
\qquad
o_t=
\operatorname{LLM}
\left(q_t,\mathcal{R}^{\mathrm{DB}}_t\right).
\tag{6}
\end{equation}

\vspace{-3mm}
Thus, conventional RAG retrieves semantically similar
records, emotion-aware RAG adds affective compatibility,
and streaming dual-brain memory continually constructs and
jointly queries separate factual and affective cognitive
structures.

\begin{figure}
    \centering
    \includegraphics[width=1\linewidth]{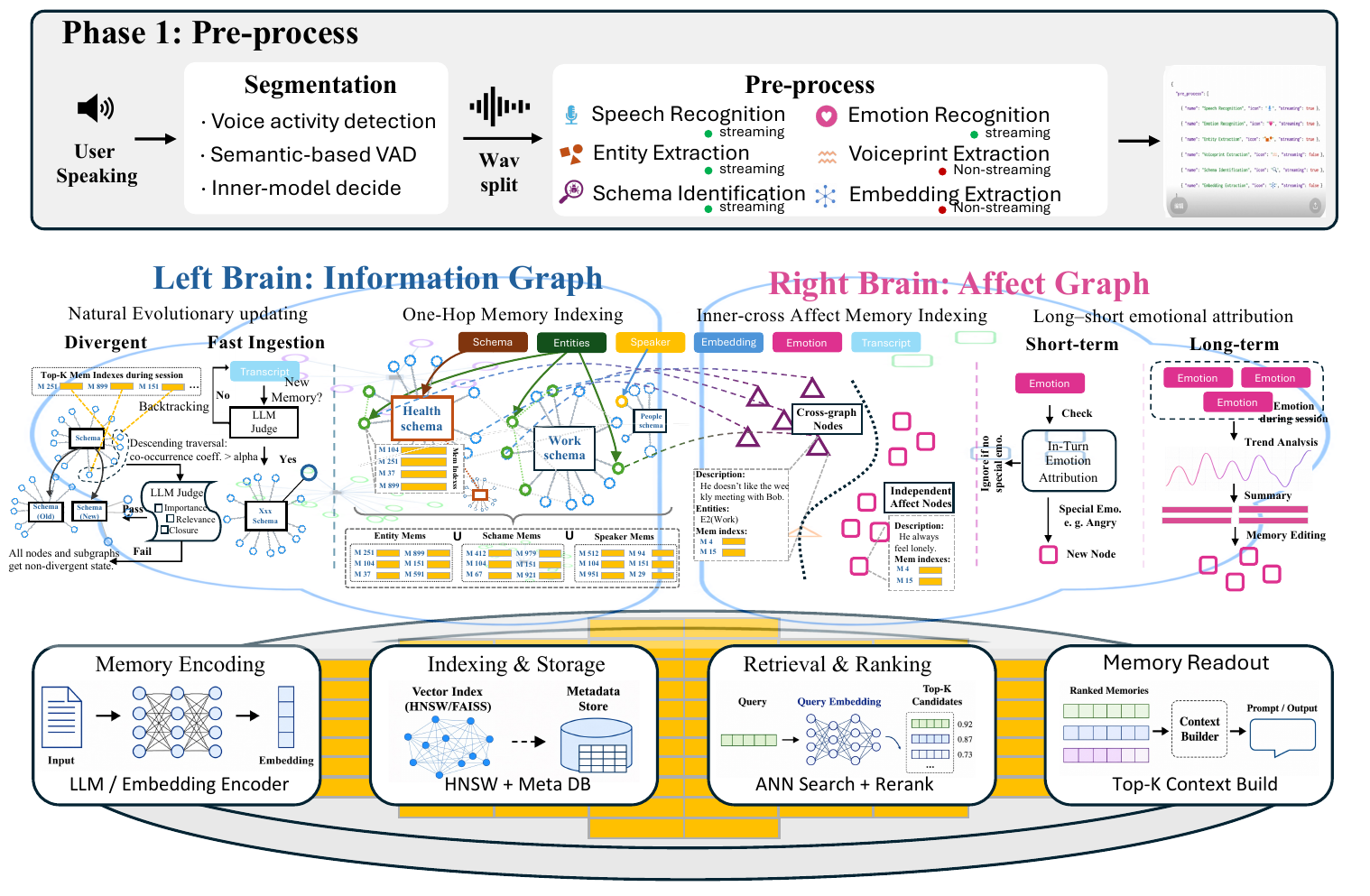}
    \vspace{-4mm}
    \caption{\textbf{The architecture of \textsc{VoiceMem}.}
    \textbf{Phase I:} Streaming preprocessing extracts speaker identity, ASR, entities, schemas, emotion, and embeddings.
    \textbf{Phase II:} The core algorithms manage the left brain via schema--entity organization and emergence, and the right brain via inner/cross-node modeling and long-/short-term emotion attribution.
    \textbf{Phase III:} Managed memory items are queried through interchangeable backend memory engines.}
    \label{fig:2-framework}
\end{figure}

\section{VoiceMem: Streaming Dual-brain Architecture}
\vspace{-1mm}
% We now instantiate streaming dual-brain archetecture as Voicemem。它必须满足以下要求：（i 有一个更高效，且有效的记忆机制。 （ii 他必须拥有同样强大的情绪和人格等副语言信息感知和记忆能力。 （iii 他的加入需要让语音对话系统感受不到延迟。

We instantiate the proposed streaming dual-brain architecture as \textsc{VoiceMem} to meet three requirements: \textbf{(i)} maximize information density under a limited memory budget; \textbf{(ii)} develop a long-term understanding of personality while learning to perceive and adapt to emotions and attitudes; and \textbf{(iii)} add no perceptible latency to real-time spoken dialogue.

\vspace{-1mm}
\subsection{Left Brain: Efficient Memory Access}
\label{sec:leftbrain}

The goal of the left brain in \textsc{VoiceMem} is to support efficient memory storage and retrieval under the tight context and latency budgets of online spoken interaction. Existing systems face two key challenges. \textbf{(i)} Retrieving up to \textbf{top-100} candidates improves coverage but overwhelms the limited context capacity of speech-language models, whereas restricting retrieval to \textbf{top-5} risks omitting relevant memories. \textbf{(ii)} Conventional memory pipelines often require \textbf{2--3 seconds} for retrieval and processing, while real-time dialogue permits only \textbf{100--200 milliseconds} of additional latency to preserve natural turn-taking.

% \subsection{左脑for 高效的信息存取}

% 对于voicemem的左脑，核心在于信息存取的高效率存取，即对于一个online 对话系统，有两个限制: i 语音模型模型能够处理的文本的能力和上下文非常有限，常用的top-100 setting in agentic memory 不仅会超出模型的接受能力，还会造成很长的处理延迟。然而，如果只用top-5或top-10，则会 即影响ii)留给记忆的检索延时很有限. 对于一个语音系统而言，通常要求<=1s的首帧延迟，而大部分记忆系统的检索流程和时间要2-3s，会破坏交互的自然性。

% 簇-实体-mem_item hypergraph indexing for 高密度信息存取。 如果想要在top-3 或5 的小范围下保证搜索精度，最核心的是检索前的有效信息密度而不是检索时的精度，eg，无关情况的同一个物体是绝对会被top-k检索调处，却起相反作用。所以我们的方式是解耦MemItem 存于mem0，而着手构建直接的轻量级上层语义索引图。

% 为了simple-but-effect，我们只放了两层结构于左脑图，即簇，一个基于语义的高级聚类中心，和下层实体，即簇语义下一个具体的人，事件或虚拟概念。二者之间的图关系为：微连接（实体-实体），表示二者实体之间有关联和宏连接（簇-簇）表示两个语义集团之间有强关联。我们去除了簇-实体连接来简化结构和避免递归查询。最终，实体和簇的定义非常简单，一个实体只拥有简单的文本description，以及自己关联的memitem index, 和弱连接实体编号。 而簇则包含自己的description，实体和强连接。 通过上层图的indexing，后续记忆检索的范围非常小，且和语义高度相关。

\textbf{Cluster--Entity--MemItem Indexing for Dense Memory Access.} \space \space When retrieval is restricted to a small budget, such as top-$5$, performance depends primarily on the semantic density of the candidate space rather than on increasingly sophisticated ranking. Multiple memory items that refer to the same entity under irrelevant contexts may dominate the top-$k$ results while providing little useful information. We therefore decouple the backend memory items stored in \textsc{Mem0} from a lightweight semantic index constructed above them. This index narrows retrieval to a compact and semantically coherent candidate set before accessing the underlying memories.

For simplicity and efficiency, the left-brain index adopts a two-level hierarchy comprising \emph{schemas} for coarse-grained semantic routing and \emph{entities} for locating concrete people, events, or concepts. We define
\[
\mathcal{G}^{L}=(\mathcal{S},\mathcal{V},\mathcal{E}), \qquad v=(d_v,\mathcal{N}^{\mathrm{micro}}_v,\mathcal{I}_v), \qquad s=(d_s,\mathcal{N}^{\mathrm{macro}}_s,\mathcal{V}_s).
\]
Each entity $v\in\mathcal{V}$ belongs to exactly one schema $s\in\mathcal{S}$, where $d_v$ and $d_s$ are textual descriptions, $\mathcal{I}_v$ indexes the associated backend memory items, and $\mathcal{V}_s$ contains the entities assigned to schema $s$. The edge set $\mathcal{E}=\mathcal{E}_{\mathrm{micro}}\cup\mathcal{E}_{\mathrm{macro}}$ supports lightweight semantic expansion: $\mathcal{N}^{\mathrm{micro}}_v$ links related entities, while $\mathcal{N}^{\mathrm{macro}}_s$ links related schemas. We encode schema membership directly rather than introducing explicit schema--entity edges, thereby avoiding recursive traversal and keeping retrieval both compact and semantically focused.

\paragraph{Retrieval.}
While the user is speaking, a streaming matcher identifies relevant schemas and entities from the partial transcript, then expands the matched entities and those contained in the matched schemas through strong and weak one-hop connections:
\begin{equation}
(\mathcal{V}_t,\mathcal{S}_t)=\operatorname{Match}(x_{\leq t},\mathcal{V},\mathcal{S}),\qquad
\mathcal{Z}_t=
\mathcal{V}_t
\cup \mathcal{V}_{\mathcal{S}_t}
\cup \mathcal{N}^{\mathrm{strong}}_1\!\left(\mathcal{V}_t\cup\mathcal{V}_{\mathcal{S}_t}\right)
\cup \mathcal{N}^{\mathrm{weak}}_1\!\left(\mathcal{V}_t\cup\mathcal{V}_{\mathcal{S}_t}\right).
\label{eq:match-expand}
\end{equation}
Here, $\mathcal{V}_t$ and $\mathcal{S}_t$ denote the matched entities and schemas, respectively; $\mathcal{V}_{\mathcal{S}_t}$ denotes the entities assigned to the matched schemas; and $\mathcal{Z}_t$ is the expanded entity set obtained through one-hop strong and weak edges. The corresponding memory items are then indexed and searched as
\begin{equation}
\mathcal{C}^{L}_t
=
\bigcup_{z\in\mathcal{Z}_t}\mathcal{I}_z,
\qquad
\mathcal{R}^{L}_t
=
\operatorname{MemSearch}(q_t,\mathcal{C}_t;K).
\label{eq:pool}
\end{equation}

\vspace{-4mm}
The backend searches only this candidate pool rather than the full memory $\mathcal{M}$. By retaining memories associated with matched and neighboring entities while sharply reducing the search space, the proposed index enables accurate and efficient top-$5$ retrieval.

\paragraph{Fast Update.}
After each turn, an asynchronous updater extracts new facts and jointly reconciles them with nearby memories through \textsc{add}, \textsc{update}, \textsc{delete}, or \textsc{keep} operations. It then updates the associated schemas, entities, and relations off the critical path; implementation details are deferred to the appendix.

% 此处实体和簇需要形式化定义

% ========== 融合前 ================
% 前向搜索：

% · 语义-实体匹配，从用户输入的信息中，会流式的提取相关的簇和实体。所有簇和实体的memitem index 会加入搜索范围。

% · 基于关系的一跳扩散：所有在前一个stage中的簇和实体会进行1步的扩散，并将新的关联到的簇和实体提取memitem index，并加入搜索范围。

% 最终，输入语句的embedding，content会使用mem0的搜索算法 实体+embedding 匹配 + top-k 在小范围中进行搜索，得到左脑的output Ileft.

% backward更新：

% · 快速的记忆插入：系统先把这一轮的 transcript 拆成一条条 fact，同时拿整段原文去做一次向量检索、捞出十条最相关的旧 mem；然后用一次大模型调用，把所有新 fact 和这十条放在一起批量判，每条给出 add / update / delete / 不变。只有 add 的才真的新存一条记忆，update 是改写旧的那条（时间戳跟着挪到本轮），delete 是把旧的删掉，后面的入图只针对新存进去的那些。接着用一次大模型调用给这批新 fact 打标注（十条一批），每条 fact  标出：它属于哪个 slot（只能从固定 slot 和子图机制已经建好的动态 slot 里选，最多两个，两个都会打进标签供检索命中，第一个当主 slot）、里面有哪些 entity 和各自的类型、entity 之间是什么关系。然后往图里写：每个 entity 先查库里有没有同名同类型的，人名只按名字查，没有才新建一个节点；接着把 entity 和这条记忆连起来，标明它在这句话里是主语还是背景，entity 之间的关系也连上边。最后把 entity 的名字转成向量，在主 slot 底下找最像的已有 entity，够像（相似度 0.65 以上）就直接挂上去复用，不够像才新开一个节点。

% 快速的查询-更新机制
% 融合后 ============= 1. 前向检索：从输入定位相关记忆
% 系统首先对用户输入进行语义-实体匹配，提取相关实体与语义簇，并定位候选记忆范围。随后基于记忆图进行一跳关系扩散，扩展相关实体与记忆节点。在局部候选空间内，结合 embedding 与实体信息进行联合检索，获取 top-k 相关记忆作为当前上下文。
% 2. 记忆判断：理解并更新新信息
% 系统将本轮交互内容拆解为结构化事实，并与已有记忆进行匹配判断，确定记忆的新增、更新或保持状态。对于变化信息，更新已有记忆状态，保证记忆内容的一致性与有效性。
% 3. 记忆入图：构建可检索结构
% 系统对新记忆进行结构化解析，提取 slot、实体及关系，并写入记忆图中建立关联。通过实体关联与语义索引优化，使新增记忆能够支持后续高效检索，实现记忆持续演化。

% ==============

% · 长期优化-涌现机制

% 从另一个视角，合适的簇的设置非常重要。越来越大的簇会导致信息密度持续降低，而目前的工作中的分裂机制会因交互系统的信息量大，次数频繁等问题导致信息过于分散，出现检索范围不足的问题。 

% 因此，我们在簇-level设计了涌现机制。总体思路为： 利用频繁的查询，反推出一个簇中真正高度相关，且足够重要的最大子图。具体实现为：给予每个实体一个分裂潜能开关。（2）在session结束后，计算每一个非所有实体都丧失分裂潜能的子图的“协同查询系数”，计算方式为： xxx。 （3）如果有某个最大子图满足协同查询系数 > a, 则进入涌现判断： 使用llm-as-a-judge，根据相关性，重要性，完整性进行判断。（4）如果全部满足，则分裂成新的簇。如果不满足，则将turn off 所有的实体的潜能。

\textbf{Cluster Emergence Mechanism.} As a cluster grows, its information density decreases, while frequent rule-based splitting may fragment related memories and reduce retrieval coverage. We therefore let coherent subclusters emerge from repeated retrieval patterns.

\noindent
\begin{minipage}[t]{0.4\textwidth}
 Let $\mathcal{Q}$ be the queries observed within a session, $A_q$ the entities activated by query $q$, and $H$ a connected subset of entities in the current cluster. We measure their query coherence by

 \vspace{-2mm}
\begin{equation}
\rho(H)=\frac{1}{|\mathcal{Q}|}\sum_{q\in \mathcal{Q}}
\frac{\left|A_q\cap H\right|}{\left|A_q\cup H\right|}.
\label{eq:rho}
\end{equation}

 \vspace{-1mm}
A high score indicates that the entities in $H$ are repeatedly retrieved together. If the largest qualifying subgraph exceeds threshold $\alpha$, an LLM judge further evaluates its relevance, importance, and completeness before promoting it to a new cluster. The whole emergence workflow is summarized in Algorithm~\ref{alg:emerge}.

\end{minipage}
\hfill 
\begin{minipage}[t]{0.58\textwidth}
\vspace{-12pt}
\footnotesize
\vspace{-3mm}
\begin{algorithm}[H]
\caption{Cluster Subgraph Emergence Pipeline}\label{alg:emerge}

\begin{algorithmic}[1]
\Require Cluster $G_c$, query set $\mathcal{Q}$, coherence threshold $\alpha$
\State $\mathcal{H}\gets \Call{ConnectedCandidates}{G_c}$
\State $H^\star\gets \arg\max_{H\in\mathcal{H}} |H|$ \hfill {\color{blue}// Filter via coherence con.}
\Statex \hspace{\algorithmicindent} subject to $\rho(H)>\alpha$
\If{$H^\star$ exists}
    \State \hfill {\color{blue}// Relevance, importance,
and completeness val.}
    \State $(r,i,c)\gets\Call{LLMJudge}{H^\star}$     
    \If{$r\land i\land c$}
        \State \Call{Promote}{$H^\star$} \hfill {\color{blue}// Do cluster emergence.}
    \Else
        \State \Call{DisablePotential}{$H^\star$} \hfill {\color{blue}//Prevent future candidate splitting}
    \EndIf
\EndIf
\end{algorithmic}
\end{algorithm}
\end{minipage}

\vspace{-2mm}
\subsection{Right Brain: Knowing the Person}
\label{sec:right}

While the left brain records what happened, the right brain captures
\emph{who the user is}. Operating in parallel, it maintains \emph{persona
memory}: stable dispositions, affective tendencies, and attitudes grounded in
specific people, events, or concepts. Such evidence spans multiple timescales,
from immediate reactions within a turn to persistent properties consolidated
across sessions.

\paragraph{Persona Graph.}
The right brain maintains two complementary types of persona nodes:
\begin{equation}
\mathcal{G}^{R}
=
\left(\mathcal{V}^{I},\mathcal{V}^{C}\right),
\qquad
v^{I}
=
\left(d_{v}^{I},\mathcal{I}_{v}^{I}\right)
\in\mathcal{V}^{I},
\qquad
v_{e}^{C}
=
\left(d_{v,e}^{C},\mathcal{I}_{v,e}^{C},\rho_{v,e}\right)
\in\mathcal{V}^{C},
\quad e\in\mathcal{V}.
\label{eq:rightgraph}
\end{equation}
Here, $d$ is a persona description and $\mathcal{I}$ indexes its supporting backend memory items. Independent nodes $v^{I}$ encode user-intrinsic properties, including enduring dispositions, behavioral regularities, and affective tendencies supported by longitudinal evidence. Cross-entity nodes $v_{e}^{C}$ instead capture context-dependent affect, with $\rho_{v,e}$ linking each node to a left-brain entity $e\in\mathcal{V}$. This distinction is fundamental: $v^{I}$ explains persistent user characteristics, whereas $v_{e}^{C}$ preserves whom or what an emotion concerns. Collapsing the two would either mistake situational reactions for stable traits or remove the real-world causes that give affect its meaning.

\paragraph{Retrieval.}
A streaming matcher jointly activates independent and cross-entity persona
nodes from the partial transcript. These matches are combined with
cross-entity nodes linked to the entities activated by the left brain:
\begin{equation}
\begin{aligned}
(\mathcal{V}^{I}_t,\mathcal{V}^{C}_t)
&=
\operatorname{Match}
\left(
x_{\leq t},
\mathcal{V}^{I},
\mathcal{V}^{C}
\right),
\qquad
\mathcal{Z}^{R}_t
=
\mathcal{V}^{I}_t
\cup
\mathcal{V}^{C}_t
\cup
\left\{
v_{e}^{C}\in\mathcal{V}^{C}:e\in\mathcal{Z}_t
\right\},
\\
\mathcal{C}^{R}_t
&=
\bigcup_{z\in\mathcal{Z}^{R}_t}\mathcal{I}_z,
\qquad
\mathcal{R}^{R}_t
=
\operatorname{MemSearch}
\left(
q_t,
\mathcal{C}^{R}_t;
K
\right).
\end{aligned}
\label{eq:rightout}
\end{equation}
Here, $\mathcal{Z}_t$ is the expanded entity set produced by the left brain
in~\eqref{eq:match-expand}. The resulting candidate pool combines persona
evidence directly implied by the conversation with attitudes grounded in the
currently relevant real-world entities, without searching the full persona
memory.

\paragraph{Short-Horizon Attribution.}
For each conversational input $x_t$, an affect estimator produces an emotion
representation $e_t=\phi(x_t)$. The pair $(x_t,e_t)$ provides immediate
evidence for adding, editing, or merging persona nodes:
\begin{equation}
e_t
=
\phi(x_t),
\qquad
\mathcal{G}^{R}_t
=
\operatorname{Modify}
\left(
\mathcal{G}^{R}_{t-1};
x_t,e_t
\right),
\qquad
t=1,\ldots,T.
\label{eq:short}
\end{equation}
Short-horizon attribution preserves the current affect together with its
situational target and cause, allowing the persona graph to adapt within the
ongoing interaction.

\paragraph{Long-Horizon Attribution.}
After each session, long-horizon attribution jointly analyzes the sequence
$(x_1,e_1),(x_2,e_2),\ldots,(x_T,e_T)$ and consolidates recurrent evidence
into stable independent persona nodes:
\begin{equation}
\mathcal{V}^{I}
\leftarrow
\operatorname{Consolidate}
\left(
\mathcal{V}^{I};
(x_1,e_1),
(x_2,e_2),
\ldots,
(x_T,e_T)
\right).
\label{eq:long}
\end{equation}
Rather than accumulating every transient state, this process identifies
persistent affective and behavioral patterns that provide a stable account of
who the user is.

\begin{figure}[!t]
    \centering
    \includegraphics[width=0.98\linewidth]{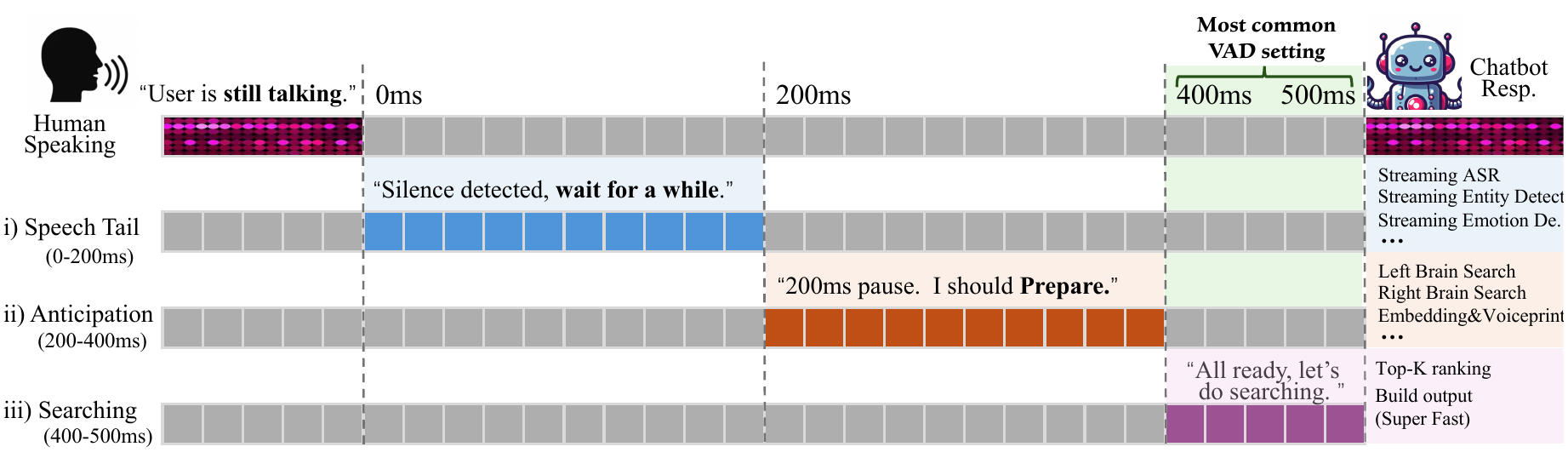}
    \caption{The four-stage streaming retrieval process, which serves as the final safeguard for reducing retrieval latency in \textsc{VoiceMem}.}
    \label{fig:streaming}
\end{figure}

\subsection{Streaming Dual-Brain Retrieval}
\label{sec:stream}

We finally describe the streaming retrieval process of \textsc{VoiceMem}, which is the core of its near-zero added latency. Four stages are introduced for this process: \textbf{\emph{listening}}, \textbf{\emph{speech tail}}, \textbf{\emph{anticipation}}, and \textbf{\emph{searching}}. During \textit{listening and (i) Speech Tail(0-200ms)}, While the user speaks, we obtain the transcript $x_{\leq t}$~\citep{shi2026qwen3,xie2026mega}, the matched entities and schemas of both brains, and the speaker identity $p_t$ in a streaming manner:
\vspace{-3mm}
\begin{equation}
\underbrace{x_{\leq t}=\operatorname{ASR}(a_{\leq t}),\;\;
(\mathcal{V}_t,\mathcal{S}_t),\;(\mathcal{V}^{I}_t,\mathcal{V}^{C}_t)=\operatorname{Match}(x_{\leq t})}_{\text{in real time}}
\;\;\Big|\;\;
\underbrace{p_t=\psi(a_{\leq t})}_{\text{(i) if delayed}}.
\label{eq:listen}
\end{equation}

\vspace{-3mm}
\textit{(ii) Anticipation. (200-400ms)} Assume a reply is coming once the silence reaches $200$\,ms. Extract the query embedding and expand the upper-layer graphs of both brains by~\eqref{eq:match-expand} and~\eqref{eq:rightout}:
\begin{equation}
q_t=\operatorname{Embed}(x_{\leq t}),
\qquad
\mathcal{Z}_t=\mathcal{G}^{L}(\mathcal{V}_t,\mathcal{S}_t),
\qquad
\mathcal{Z}^{R}_t=\mathcal{G}^{R}(\mathcal{V}^{I}_t,\mathcal{V}^{C}_t,\mathcal{Z}_t).
\label{eq:prep}
\end{equation}

\vspace{-2mm}
\textit{(iii) Searching.(400-500ms)}
Only the backend search is left. Both brains are searched and merged:
\begin{equation}
\mathcal{R}^{L}_t=\operatorname{MemSearch}(q_t,\mathcal{C}^{L}_t;K),
\space
\mathcal{R}^{R}_t=\operatorname{MemSearch}(q_t,\mathcal{C}^{R}_t;K),
\space
R_t=\operatorname{Prompt}(\mathcal{R}^{L}_t,\mathcal{R}^{R}_t).
\label{eq:query}
\end{equation}
A VAD threshold is commonly set to $500$\,ms, leaving a $400$\,ms window before the reply must start; \textsc{VoiceMem} meets this budget with margin, as the dense dual-brain retrieval itself costs only $134$\,ms.
\subsection{What Was the Song Playing at the Café Yesterday?}

\textsc{VoiceMem} extends memory beyond text to audio, supporting multi-speaker discrimination, paralinguistic analysis, and environmental sound memory. When enabled, agents selectively retain audio as \textbf{\underline{speaker voiceprints}}, \textbf{\underline{acoustic embeddings}}, or \textbf{\underline{raw waveforms}}, and attach them as multimodal nodes to the corresponding entity nodes.

% \subsection{右脑for 情感和以人为中心的了解}

% 一个好的助手除了记住所有事情，还需要了解自己的主人。具有对主人情绪和人格的了解和高智商同样重要，这就是我们设立与左脑平行的右脑的初衷。情绪存储面临两个核心挑战：1. 复杂的内/外情感 和 2. 复杂的深-浅层人格记忆， 分别对应着voicemem的结构和存取机制。

% \paragraph inner-cross 记忆节点

% 两种节点，独立节点和cross节点。 读取时分别会触发左脑物体关联的情感节点，右脑独立节点，和类似情绪的节点。 右脑的簇固定，去掉了所有的关系。每个实体的description并非用于分裂，而是直接进入信息output。

% long-short 情感归因。 每个turn /  session 之后基于情感流归因。

% \subsection 流式dual brain记忆检索
% 最后，我们最后介绍voicemem的流式检索过程。我们将一次对话回复前的时间分为四个时间区域：聆听时段，空白时段，预备时段，记忆查询时段。我们充分利用每格stage的时间。 细节为： 聆听时段： for 每个输入的句子做 实时语音识别，实时实体识别，实时语义簇匹配，和声纹提取。空白时段： 为0-200ms，for 最后一句的后处理时间，以及声纹提取/匹配完成。预备时段：200ms-400ms，这一阶段代表上一段语音已经停止200ms。即可能是一次需要应答。此时则开始语义embedding提取，以及上层图的扩散查询indexing。记忆查询时段：400-500ms：for 最终的mem0查询，左右脑信息合并以及prompt构建。 由此，通常的vad值设为500ms，而voicemem亦在同样延时下完成了高密度记忆的查询。

\section{Infrastructure: Model Training, Validation and Deployment}
In this section, we provide implementation details of \textsc{VoiceMem} in real-world settings, focusing on two aspects: i) \emph{model training}, where we convert the Qwen2.5-Omni~\citep{qwen25omni}, Qwen3-Omni~\citep{qwen3omniblog}, and Step-Audio2-Mini~\citep{stepaudio2} model families--originally  designed for the speech-input/text-output--into memory-augmented speech language models through online black-box on-policy distillation, yielding, to our knowledge, the first speech language models with explicit memory access; and ii) \emph{deployment}, where we design an evolvable two-layer system architecture for \textsc{VoiceMem} that decouples the memory layer from the underlying search engine.

\begin{figure}[t]
    \centering
    \includegraphics[width=1\linewidth]{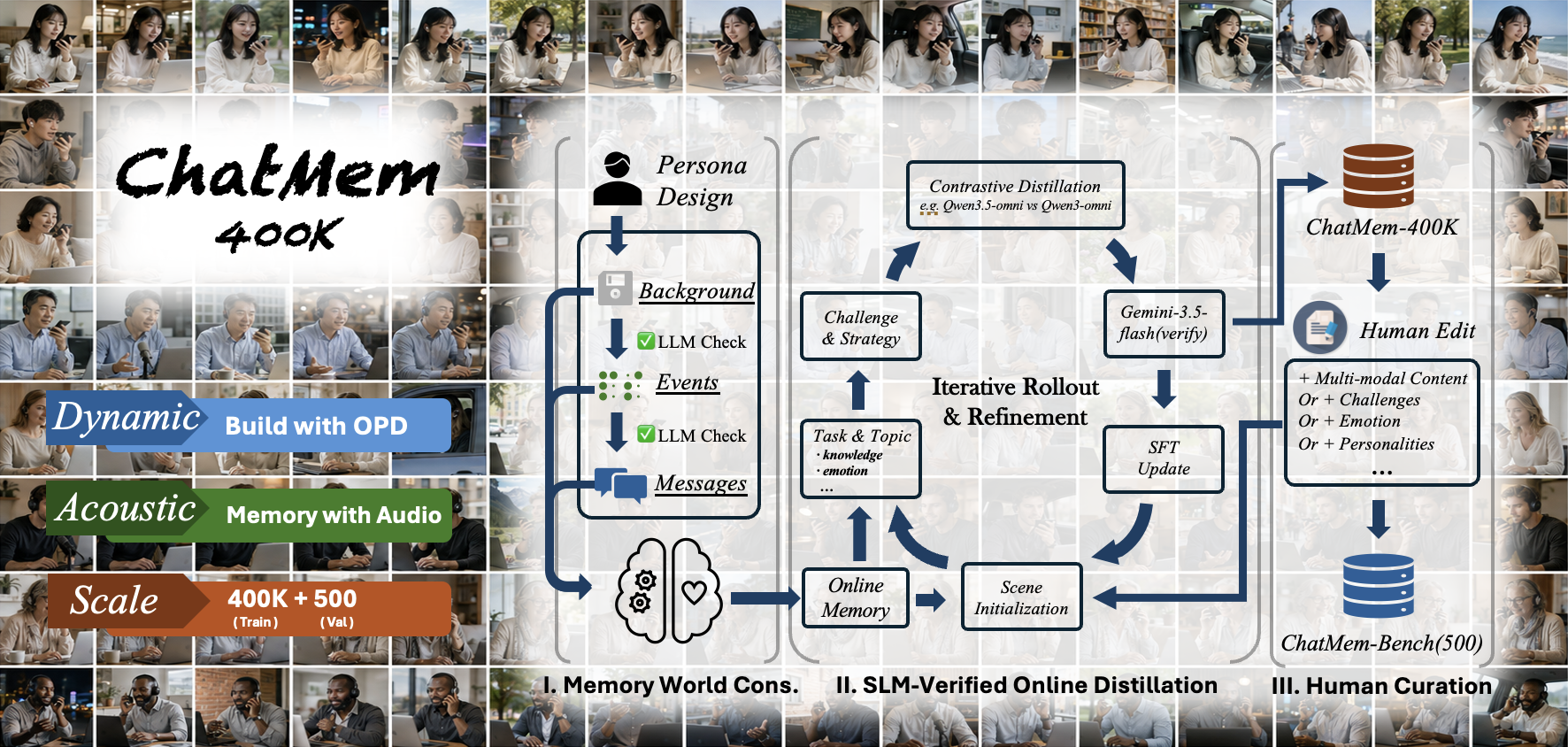}
    \caption{\textbf{Training and validation pipeline for \textsc{VoiceMem}.}
We construct \textsc{ChatMem-400K} through three stages: memory-world construction, SLM-verified online on-policy distillation, and human curation, covering dynamic user histories, acoustic inputs, and large-scale memory-dependent conversations.
A human-curated subset further forms \textsc{ChatMem-Bench} for evaluating information recall, persona understanding, affective attribution, and paralinguistic/environmental reasoning.}   
   \label{fig:training-pipeline}
\end{figure}

\subsection{Black-box OPD Training}
\vspace{-1mm}
To maximize model performance, we employ proprietary models as teachers and adopt online distillation to mitigate catastrophic forgetting during adaptation. Specifically, Qwen2.5-Omni and Qwen3-Omni are distilled from Qwen3.5-Omni~\citep{qwen35omniblog}, while Step-Audio2-Mini is distilled from Step-Audio2~\citep{stepaudio2}. The Training pipeline consists of four stages:

\label{sec:data-construction}
\noindent\textbf{Stage I: Memory World Construction.}
We first construct a coherent long-term memory world for each synthetic user. Starting from a core persona, we progressively instantiate background information, life events, messages, and online memories:
$\mathsf{Persona}\rightarrow\mathsf{Background}\rightarrow\mathsf{Events}\rightarrow\mathsf{Messages}\rightarrow\mathsf{Memory}$.
This yields temporally consistent user histories for downstream dialogue generation.

\vspace{-1mm}
\noindent\textbf{Stage II: SLM-Verified Online Distillation.}
We generate memory-dependent conversations through an iterative pipeline: i) \emph{Task \& Topic} samples knowledge-, emotion-, or persona-oriented goals; ii) \emph{Scene Initialization} constructs the current context; iii) \emph{Challenge \& Strategy} introduces recall, reference resolution, contradiction, and multi-memory reasoning; iv) \emph{Contrastive Distillation} compares responses with different memory-use behaviors; v) \emph{Verification} filters for necessity, faithfulness, and quality; and vi) \emph{SFT Update} updates the generator for the next iteration. Repeating this loop yields \textsc{ChatMem-400K}, a large-scale corpus of memory-dependent conversations.

\vspace{-1mm}
\noindent\textbf{Stage III: Human Curation.}
Human annotators further refine the corpus and construct harder samples with richer emotional and personality dynamics and multimodal inputs such as acoustic questions. These examples augment training, while a challenging subset is curated into \textsc{ChatMem-Bench}.

\vspace{-1mm}
\noindent\textbf{Stage IV: Validation.}\label{sec:chatmem-bench-method}
% 我们使用ChatMem-Bench 作为最终模型的评价benchmark之一，for 目前没有完全匹配的场球语音助手的benchmark，主要包含四个维度：信息，人格，情感归因和副语言&环境和28个子类问题。 由于数据构造耗费大量人力，我们会在following的单独report中详细介绍。
We use \textsc{ChatMem-Bench} as one of the primary benchmarks for evaluating the final models, as existing benchmarks do not fully capture the requirements of real-world voice assistants. 
It covers four dimensions:
\underline{\textbf{\textit{\{Information, Persona, Affective Attribution, Paralinguist-}}}
\underline{\textbf{\textit{ics \& Environment\}}}}, across 14 fine-grained question categories. As its construction requires substantial human effort, we defer a detailed introduction of the benchmark and annotation pipeline to a separate follow-up technical report.

\vspace{-1mm}
\subsection{Decoupled Upper-level Routing--Lower-level Engine Architecture}

% 底层的memory正在持续进化，参数化和latent的memory也正在探索。我们不希望实时的记忆架构和更先进的记忆机制独立发展，for 无法及时更新和兼容最新算法。因此，我们用了额外的大量时间来重新解耦架构。 voicemem的探索更注重上层路由，并行双脑架构以及流式的推理机制i，而第三章中的memsearch函数所代表的底层引擎是完全解耦的。暂时的，我们使用mem0作为底层引擎for它仍然通用且sota性能。
The underlying memory engine is evolving rapidly, \space with parameterized and latent memory mechanisms also emerging as promising directions. \space We therefore avoid tightly coupling the real-time memory architecture with any particular backend, which would otherwise hinder timely adoption of newer memory algorithms. To this end, we substantially redesigned \textsc{VoiceMem} around a decoupled architecture: our focus lies on upper-level routing, the parallel dual-brain organization, and streaming reasoning, while the underlying engine abstracted by $\operatorname{MemSearch}$ in Section~3 remains fully interchangeable. In our current implementation, we use \textsc{Mem0} as the backend due to its generality and strong state-of-the-art performance.

\makeatletter
% -- 本节所需定义（已存在则跳过，不会冲突）--
\providecommand{\ph}[1]{\textcolor{red}{#1}}
\definecolor{headerblue}{RGB}{234,240,250}
\providecolor{srcblue}{RGB}{18,62,138}
\providecolor{estorange}{RGB}{201,110,17}
\providecommand{\src}[1]{\textcolor{srcblue}{#1}}
\providecommand{\est}[1]{\textcolor{estorange}{#1}}
\expandafter\ifx\csname NC@rewrite@C\endcsname\relax
  \newcolumntype{C}[1]{>{\centering\arraybackslash}p{#1}}
\fi
% -- 排版：图文间距 / 页面空洞 / 浮动体分布 --
\setlength{\textfloatsep}{8pt plus 2pt minus 2pt}
\setlength{\intextsep}{8pt plus 2pt minus 2pt}
\setlength{\floatsep}{8pt plus 2pt minus 2pt}
\setlength{\abovecaptionskip}{5pt}
\setlength{\belowcaptionskip}{2pt}
\renewcommand{\topfraction}{0.94}
\renewcommand{\bottomfraction}{0.65}
\renewcommand{\textfraction}{0.05}
\renewcommand{\floatpagefraction}{0.88}
\setcounter{topnumber}{4}
\setcounter{bottomnumber}{3}
\setcounter{totalnumber}{6}
%% no running head in this section: the rule and the section name cost two
%% text lines on every page. Move \pagestyle{plain} to the preamble to drop
%% them document-wide.
\pagestyle{plain}\thispagestyle{plain}
%% Tighter leading. article's default at 10pt is 12pt baselineskip; 0.94 puts it
%% near 11.3pt. Move this to the preamble to apply document-wide.
\linespread{0.94}\selectfont
%% Tighter headings. titlesec is not loaded by the class, so we redefine the
%% two levels through \@startsection directly (we are inside \makeatletter).
\renewcommand\section{\@startsection{section}{1}{\z@}%
  {-10pt \@plus -2pt \@minus -2pt}{4pt}{\normalfont\Large\bfseries}}
\renewcommand\subsection{\@startsection{subsection}{2}{\z@}%
  {-7pt \@plus -1pt \@minus -2pt}{2.5pt}{\normalfont\large\bfseries}}
%% run-in \paragraph: the default 3.25ex before-skip leaves the three Settings
%% blocks floating apart from one another.
\renewcommand\paragraph{\@startsection{paragraph}{4}{\z@}%
  {1.1ex \@plus .3ex \@minus .2ex}{-0.55em}{\normalfont\normalsize\bfseries}}
\raggedbottom
\setlength{\@fptop}{0pt}
\setlength{\@fpsep}{16pt}
\setlength{\@fpbot}{0pt plus 1fil}
\setlength{\@dblfptop}{0pt}
\setlength{\@dblfpsep}{16pt}
\setlength{\@dblfpbot}{0pt plus 1fil}
\makeatother
% -- 浮动栅栏：小节边界处强制排空队列，避免图表漂到后面几页 --
\providecommand{\floatbarrier}{\par\bigskip\penalty-200\relax
  \@ifundefined{FloatBarrier}{\clearpage}{\FloatBarrier}}

\definecolor{hltop1}{RGB}{140,190,238}
\definecolor{hltop2}{RGB}{214,233,250}
\newcommand{\hlone}[1]{\cellcolor{hltop1}#1}
\newcommand{\hltwo}[1]{\cellcolor{hltop2}#1}
%%%%%%%%%%%%%%%%%%%%%%%%%%%%%%%%%%%%%%%%%%%%%%%%%%%%%%%%%%%%%%%%%%%%%%%%%%%%%

% -- 排名荧光笔高亮：深蓝=第一名，浅蓝=第二名 --
% -- 排名荧光笔高亮：深蓝=第一名，浅蓝=第二名 --
\definecolor{hltop1}{RGB}{140,190,238}
\definecolor{hltop2}{RGB}{214,233,250}
\setlength{\fboxsep}{1pt}

\begin{table}[!t]
\centering
% ==================== Table 1 ====================
\caption{\textbf{Factual memory results.} LLM-judge score (\%, $\uparrow$).
$\ddagger$: responses generated by our fine-tuned {Qwen3.6}; all other
rows use GPT-4o-mini as the response model.}
\label{tab:left}
\vspace{-0.1em}
{\fontsize{8.5pt}{10pt}\selectfont
\setlength{\tabcolsep}{2.59pt}
\renewcommand{\arraystretch}{0.92}
\begin{tabular}{@{}l *{11}{C{0.92cm}} C{0.92cm}@{}}
\toprule
\multirow{2}{*}{\textbf{Method}}
  & \multicolumn{4}{c}{\textbf{LoCoMo}}
  & \multicolumn{4}{c}{\textbf{LongMemEval}}
  & \multicolumn{3}{c}{\textbf{Memora}}
  & \multirow{2}{*}{\textbf{Avg.}} \\
\cmidrule(lr){2-5}\cmidrule(lr){6-9}\cmidrule(lr){10-12}
  & Single & Multi & Temp. & Open
  & Extr. & Multi & Upd. & Temp.
  & Rem. & Rec. & Rea. & \\
\midrule
\multicolumn{13}{@{}l}{\cellcolor{headerblue}\textit{\textbf{Reference}}} \\
Full-Context
  & 75.87 & 72.27 & 70.29 & 57.45
  & 45.45 & 73.92 & 78.21 & 31.58
  & \hlone{85.54} & 52.10 & 22.67 & 60.49 \\
\midrule
\multicolumn{13}{@{}l}{\cellcolor{headerblue}\textit{\textbf{Memory Engines}}} \\
Mem0
  & 67.13 & 51.15 & 55.51 & 72.93
  & 62.82 & 46.21 & 70.12 & 40.15
  & 40.42 & 52.58 & 16.00 & 52.27 \\
Zep
  & 66.91& 53.70 & 60.29 & 70.82
  & 47.46 & \hltwo{74.44} & 54.10 & 34.62
  & 33.36 & 37.38 & 3.50 & 48.78 \\
LangMem
  & 62.23 & 47.92 & 43.43 & 71.12
  & 55.13 & 20.30 & 66.67 & 15.79
  & 68.16 & 48.88 & \hltwo{30.00} & 48.15 \\
\midrule
\multicolumn{13}{@{}l}{\cellcolor{headerblue}\textit{\textbf{Agent Memory}}} \\
A-MEM
  & 49.79 & 49.85 & 54.05 & 49.91
  & 64.62 & 48.87 & 64.11 & 47.36
  & 68.82 & 35.04 & 2.00 & 48.58 \\
MemoryOS
  & 69.35 & 62.30 & 44.72 & 44.29
  & 58.10 & 31.06 & 48.72 & 32.33
  & 15.24 & 38.86 & 1.50 & 40.59 \\
MemOS
  & 81.09 & 67.49 & 75.18 & 55.90
  & \hlone{86.77} & 70.70 & 74.33 & 77.49
  & 51.84 & \hlone{62.64} & 20.66 & 65.83 \\
\midrule
\multicolumn{13}{@{}l}{\cellcolor{headerblue}\textit{\textbf{Personal Memory}}} \\
MemoryBank
  & 21.25 & 18.40 & 31.50 & 26.28
  & 23.33 & 26.00 & 14.00 & 20.00
  & 22.22 & 41.88 & 1.42 & 22.39 \\
EverMemOS
  & 90.12 & 87.43 & 85.16 & 69.79
  & 82.71 & 73.68 & \hlone{89.74} & 77.44
  & 17.21 & 33.27 & 16.67 & 65.75 \\
Emotional RAG
  & 19.62 & 25.53 & 34.89 & 13.54
  & 40.00 & 55.00 & 70.00 & 65.00
  & 51.16 & 39.38 & 14.75 & 38.99 \\
\midrule
\multicolumn{13}{@{}l}{\cellcolor{headerblue}\textit{\textbf{Ours}}} \\
\textbf{VoiceMem}
  & \hltwo{94.60} & \hltwo{91.60} & \hltwo{89.80} & \hltwo{85.60}
  & \hltwo{83.20} & \hlone{75.00} & 77.50 & \hlone{95.00}
  & 66.80 & 50.50 & \hlone{30.70}
 & \hlone{76.39} \\
\textbf{VoiceMem}$^{\ddagger}$
  & \hlone{95.15} & \hlone{92.40} & \hlone{91.35} & \hlone{86.08}
  & 79.64 & 70.17 & \hltwo{79.24} & \hltwo{89.63} & \hltwo{69.10}
  & \hltwo{54.77} & 24.61 & \hltwo{75.65}\\
\bottomrule
\end{tabular}

% ==================== Space between tables ====================
\vspace{-0.3em}
% ==================== Table 2 ====================
\caption{\textbf{Persona memory results.} LLM-judge score (\%, $\uparrow$).}
\label{tab:right}
\vspace{-0.1em}
{\fontsize{8.5pt}{10pt}\selectfont
\setlength{\tabcolsep}{2pt}
\renewcommand{\arraystretch}{0.92}
\begin{tabular}{@{}l *{11}{C{0.95cm}} C{0.95cm}@{}}
\toprule
\multirow{2}{*}{\textbf{Method}}
  & \multicolumn{4}{c}{\textbf{ES-MemEval}}
  & \multicolumn{5}{c}{\textbf{PersonaMem}}
  & \multicolumn{2}{c}{\textbf{PersonaLens}}
  & \multirow{2}{*}{\textbf{Avg.}} \\
\cmidrule(lr){2-5}\cmidrule(lr){6-10}\cmidrule(lr){11-12}
  & IE & TR & CD & UM
  & Rec. & Gen. & Evol. & Reason & Reco.
  & Per. & TCR & \\
\midrule
\multicolumn{13}{@{}l}{\cellcolor{headerblue}\textit{\textbf{Reference}}} \\
Full-Context
  & 20.20 & 19.60 & 12.10 & 21.70
  & 41.00 & 46.00 & \hltwo{68.00} & 77.00 & 37.00
  & 75.00 & 91.08 & 46.24 \\
\midrule
\multicolumn{13}{@{}l}{\cellcolor{headerblue}\textit{\textbf{Memory Engines}}} \\
Mem0
  & 75.10 & 56.70 & 72.30 & 64.20
  & 32.13 & 57.89 & 54.68 & 80.81 & 52.73
  & 82.25 & \hltwo{92.39} & 65.56 \\
Zep
  & 66.80 & 42.30 & 64.20 & 58.00
  & 35.30 & 42.10 & 34.50 & 50.50 & 36.40
  & 58.50 & 82.74 & 51.94 \\
LangMem
  & 18.80 & 15.00 & 20.40 & 22.70
  & 31.29 & 8.77 & 53.24 & 81.82 & 40.00
  & 77.45 & 88.99 & 41.68\\
\midrule
\multicolumn{13}{@{}l}{\cellcolor{headerblue}\textit{\textbf{Agent Memory}}} \\
A-MEM
  & 67.80 & 58.30 & 67.00 & 56.70
  & 63.01 & 57.89 & 54.68 & \hlone{85.86} & 69.09
  & 62.50 & 83.26 & 66.01 \\
MemoryOS
  & 43.90 & 30.50 & 41.00 & 41.50
  & 47.06 & 56.14 & 32.37 & 48.48 & 47.27
  & 55.75 & 82.01 & 47.82 \\
MemOS
  & 77.70 & 64.60 & 66.30 & 73.50
  & 72.72 & 56.14 & 58.27 & 78.79 & \hlone{72.72}
  & 82.75 & 91.47 & 72.27\\
\midrule
\multicolumn{13}{@{}l}{\cellcolor{headerblue}\textit{\textbf{Personal Memory}}} \\
MemoryBank
  & 43.00 & 35.20 & 42.70 & 38.70
  & 35.29 & 49.12 & 51.08 & 56.57 & 40.00
  & 64.50 & 80.66 & 48.80 \\
EverMemOS
  & 66.80 & 50.70 & 59.60 & 45.90
  & 47.06 & 47.37 & 41.01 & 43.43 & 56.36
  & 63.25 & 79.91 & 54.67 \\
Emotional RAG
  & \hltwo{82.30} & \hltwo{67.10} & \hlone{77.30} & \hltwo{73.70}
  & 52.94 & 61.40 & 44.60 & 58.59 & 43.64
  & 64.25 & 82.00 & 64.35 \\
\midrule
\multicolumn{13}{@{}l}{\cellcolor{headerblue}\textit{\textbf{Ours}}} \\
\textbf{VoiceMem}
  & \hlone{82.80} & \hlone{67.30} & 69.10 & \hlone{74.00}
  & \hlone{76.47} & \hltwo{64.91} & 51.08 & 83.84 & \hltwo{70.91}
  & \hltwo{83.75} & 91.63 & \hltwo{74.16} \\
\textbf{VoiceMem}$^{\ddagger}$
  & 81.20 & 64.70 & \hltwo{73.50} & 71.90
  & \hltwo{75.19} & \hlone{70.17} & \hlone{73.38} & \hltwo{83.94} & 69.10
  & \hlone{85.50} & \hlone{93.57} & \hlone{76.56}\\
\bottomrule
\end{tabular}
}}
\end{table}

\section{Experiments}
\label{sec:experiments}

In this section, we conduct a series of experiments to answer four questions:

\begin{tcolorbox}[
colback=gray!8,
colframe=black,
boxrule=0.6pt,
arc=3mm,
left=2mm,
right=2mm,
top=1.5mm,
bottom=1.5mm
]
\textbf{(RQ1) Does VoiceMem retrieve accurately across modalities?}
\emph{\S\ref{sec:main}, \S\ref{sec:voice}}
\vspace{1mm}

\textbf{(RQ2) Is that accuracy cheap and fast enough for a live voice loop, and
what makes it so?}
\emph{\S\ref{sec:abl}}
\vspace{1mm}

\textbf{(RQ3) Does every component earn its place, and what memory structure do
they produce?}
\emph{\S\ref{sec:abl}}
\vspace{1mm}

\textbf{(RQ4) Does the index transfer across different underlying memory stores?}
\emph{\S\ref{sec:abl}}
\end{tcolorbox}

%% ==================================================================
\subsection{Settings}
\label{sec:settings}

\paragraph{Benchmarks.}
We evaluate on three families.
\emph{(i) Information memory}: LoCoMo~\citep{locomo},
LongMemEval (S)~\citep{longmemeval}, and Memora~\citep{memora}.
\emph{(ii) Persona memory}: ES-MemEval~\citep{esmemeval},
PersonaMem~\citep{personamem}, and PersonaLens~\citep{personalens}.
\emph{(iii) Voice-grounded memory}: ChatMem-Bench, with $316$ questions
over $53$ hours of audio in four ability groups and fourteen sub-categories.

\paragraph{Baselines.}
Ten systems in four groups.
\emph{(i) Reference}: Full-Context.
\emph{(ii) Memory engines}: Mem0~\citep{mem0}, Zep~\citep{zep},
LangMem~\citep{langmem}.
\emph{(iii) Agent memory}: A-MEM~\citep{a-mem}, MemoryOS~\citep{memoryos},
MemOS~\citep{memos}.
\emph{(iv) Personal memory}: MemoryBank~\citep{memorybank},
EverMemOS~\citep{evermemos}, Emotional RAG~\citep{emotionalrag}.

\paragraph{Implementation details.}
All baseline systems use GPT-4o-mini~\citep{gpt4omini} as the backbone LLM and
text-embedding-3-small~\citep{textembedding3small} as the embedding model, with
temperature $0$. VoiceMem is evaluated under six retrieval settings,
$K\in\{1,3,5,10,30,100\}$ with $K{=}5$ as the default operating point. 
%% ==================================================================
\subsection{Main Results}
\label{sec:main}

\textbf{For [RQ1.1] on text conversations}, \textsc{VoiceMem} retrieves dense memory on both information and persona. As presented in Tab.~\ref{tab:left}, our method leads seven of the eleven information sub-categories. It averages 76.39, beating Mem0, its own storage backend, by +24.12 and full-context processing, which sees the entire history, by +15.90. The margin is widest when several memories must arrive together (e.g., temporal reasoning, +54.9) and narrowest on update tracking (+7.4), which a single most-recent memory already answers. As presented in Tab.~\ref{tab:right}, our method reaches 74.16 on eleven persona
sub-categories with GPT-4o-mini and 76.56 with our fine-tuned response model,
passing the strongest baseline, MemOS, by +1.89. The reference inverts on ES-MemEval, where full context scores 12.10 on conflict detection and 21.70 on user modeling, against 69.10 and 74.00 for ours, i.e., the evidence sits in its input and it still cannot attribute it. Notably, both tables are produced at a retrieval budget an order of magnitude below the top-$100$ setting conventional in text-agent memory, which indicates that the gains come from a denser candidate pool rather than a larger one.

%% ==================================================================
\subsection{ChatMem-Bench: Long-Horizon Audio Assistant Memory}
\label{sec:voice}

\begin{table}[!t]
\centering
\caption{\textbf{ChatMem-Bench results.} LLM-judge score (\%, $\uparrow$).}
\label{tab:voice}
\vspace{-0.1em}
{\fontsize{7.8pt}{10pt}\selectfont
\setlength{\tabcolsep}{0.22pt}
\renewcommand{\arraystretch}{0.85}
\begin{tabular}{@{}l *{14}{C{0.88cm}} C{0.88cm}@{}}
\toprule
\multirow{2}{*}{\textbf{Method}}
  & \multicolumn{4}{c}{\textbf{Information}}
  & \multicolumn{3}{c}{\textbf{Persona}}
  & \multicolumn{3}{c}{\textbf{Affective Attr.}}
  & \multicolumn{4}{c}{\textbf{Paraling.\&Env.}}
  & \multirow{2}{*}{\textbf{Avg.}} \\
\cmidrule(lr){2-5}\cmidrule(lr){6-8}\cmidrule(lr){9-11}\cmidrule(lr){12-15}
  & Upd. & Temp. & Synt. & Abst.
  & Prof. & Pref. & Impl.
  & Attr. & Agd. & Phras.
  & Bsr. & Asi. & Mul. & Asa. & \\
\midrule
\multicolumn{16}{@{}l}{\cellcolor{headerblue}\textit{\textbf{Reference}}} \\
Full-Context
  & 62.07 & \hltwo{42.11} & 68.42 & 57.89 & 61.11 & 47.83 & 41.17
  & 16.67 & 57.14 & 37.78 & 11.76 & 22.58 & 26.92 & 90.00 & 45.96 \\
\midrule
\multicolumn{16}{@{}l}{\cellcolor{headerblue}\textit{\textbf{Memory Engines}}} \\
Mem0
  & 55.17 & 31.57 & 83.21 & 68.42 & 66.67 & 78.26 & 52.94
  & 16.67 & 61.90 & 46.67 & 11.76 & 3.23 & 11.54 & 93.00 & 48.64 \\
Zep
  & 37.93 & 15.78 & 63.16 & 63.16 & 50.00 & 69.57 & 35.29
  & 8.33 & 57.14 & 33.33 & 5.88 & 9.68 & 7.69 & 91.50 & 39.17 \\
LangMem
  & 58.62 & 21.05 & 73.68 & 78.95 & 61.11 & 73.91 & 47.05
  & 41.67 & 47.62 & 42.22 & 11.76 & 12.90 & 15.38 & 93.00 & 48.49 \\
\midrule
\multicolumn{16}{@{}l}{\cellcolor{headerblue}\textit{\textbf{Agent Memory}}} \\
A-MEM
  & 41.38 & 10.53 & 68.42 & 73.68 & 38.89 & 73.91 & 41.17
  & 25.00 & 52.38 & 44.44 & 23.53 & 9.68 & 3.85 & 93.65 & 42.89 \\
MemoryOS
  & 51.71 & 36.84 & \hlone{89.47} & 73.68 & 44.44 & \hltwo{82.61} & 35.29
  & 25.00 & 52.38 & 35.55 & 5.88 & 19.35 & 15.38 & \hltwo{95.59} & 47.37 \\
MemOS
  & 72.41 & 31.57 & 52.63 & 84.21 & \hlone{77.78} & 78.26 & 47.05
  & 50.00 & \hltwo{71.43} & 46.67 & 17.65 & 12.90 & 19.23 & 93.46 & 53.95 \\
\midrule
\multicolumn{16}{@{}l}{\cellcolor{headerblue}\textit{\textbf{Personal Memory}}} \\
MemoryBank
  & 41.38 & 5.26 & 47.37 & 63.16 & 38.89 & 52.17 & 29.41
  & 8.33 & 42.86 & 24.44 & 23.53 & 6.45 & 7.69 & 94.49 & 34.67 \\
EverMemOS
  & 44.83 & 26.32 & 63.16 & \hlone{94.74} & 50.00 & 69.57 & 52.94
  & 33.33 & 66.67 & 37.78 & 11.76 & 9.68 & 23.08 & 94.41 & 48.45 \\
Emotional RAG
  & 34.48 & 10.53 & 57.89 & 73.68 & 44.44 & 60.87 & \hltwo{58.82}
  & \hltwo{58.33} & 52.38 & 42.22 & 11.76 & 3.23 & 3.85 & 94.55 & 43.36 \\
\midrule
\multicolumn{16}{@{}l}{\cellcolor{headerblue}\textit{\textbf{Ours}}} \\
\textbf{VoiceMem}
  & \hltwo{79.31} & \hlone{47.37} & \hltwo{83.68} & \hltwo{89.47} & \hltwo{72.22} & \hlone{86.96} & \hlone{64.71}
  & \hlone{66.67} & \hlone{76.19} & \hlone{51.11} & \hltwo{47.06} & \hltwo{45.16}
  & \hlone{53.84} & \hlone{98.50} & \hlone{68.73} \\
\textbf{VoiceMem}$^{\ddagger}$
  & \hlone{86.21} & \hltwo{42.11} & 83.16 & \hltwo{89.47} & 66.67 & \hlone{86.96} & \hltwo{58.82}
  & \hltwo{58.33} & 66.67 & \hltwo{48.89} & \hlone{52.94} & \hlone{51.61} & \hltwo{50.00} & 95.56 & \hltwo{66.96} \\
\bottomrule
\end{tabular}
}
\end{table}

As briefly introduced in Sec.~\ref{sec:chatmem-bench-method}, \underline{\textbf{\textsc{ChatMem-Bench} evaluates whether a memory system can}}  \underline{\textbf{consolidate long-term speech interaction, across complex environments, affect, and informat-}}  \underline{\textbf{ion into a good voice assistant}}. It contains \textbf{316} questions drawn from \textbf{15,314} turns and 53 hours of dialogue, and every session is \textbf{\textit{curated by humans}}, from choosing the topic to constructing the memory world and designing the challenge. As shown in Tab.~3, the benchmark has four dimensions with 14 fine-grained categories. From left to right, Information covers \textit{update tracking}, \textit{temporal reasoning}, \textit{synthesis}, and \textit{abstention}; Persona covers \textit{profile constraints}, \textit{preference-aligned recommendation}, and \textit{implicit-need inference}; Affective Attribution covers \textit{emotion--target attribution}, \textit{attachment-guided decisions}, and \textit{affect-aware phrasing}; Paralinguistics \& Environment covers \textit{background-sound recall}, \textit{acoustic-scene inference}, \textit{multi-user conversational memory}, and \textit{acoustic style adaptation}.

\textbf{For [RQ1.2] on long multi-turn audio,} as shown in Tab.~\ref{tab:voice}, \textsc{VoiceMem} leads on 11 of the 14 categories, and the gap is widest on Paralinguistics \& Environment, where a transcript carries no evidence at all: every text system lands between 3.23 and 26.92 on the three acoustic categories, while ours reaches 45.16 to 53.84. Affective Attribution shows the opposite, i.e., text baselines stay competitive because word choice already carries most of the affective signal, and the audio increment appears instead on attachment-guided decisions and affect-aware phrasing. 

%% 原 fig1_streaming.pdf（二维 deployment frontier）已删除：
%% 它和 §5.4 的三维前沿图讲的是同一件事，两张 caption 都是 "deployment frontier
%% on LoCoMo"，重复。附录 appendix_deployment.tex 里 "the full per-system table
%% behind Figure~\ref{fig:stream}" 这句现在会悬空，需要改成指向 §5.4 的
%% Figure~\ref{fig:frontier3d}。

%% ==================================================================
%%  §5.4 Ablation and Analysis
%%  段落顺序：RQ2（前沿图、K-sweep）→ RQ3（柱状图、散点图）→ RQ4（迁移表）
%%  每段只分析一张图/表。
%%
%%  需要的图（放 visuals/，覆盖同名旧文件）：
%%    visuals/vm_combined.png   —— 左 3D 前沿 + 右 K-sweep，一张图两个 caption
%%    fig4_component.png        —— 四数据集组件消融柱状图
%%    fig3_dualbrain.png        —— 双脑散点
%% ==================================================================
\subsection{Ablation and Analysis}
\label{sec:abl}

\begin{figure}
    \centering
    \includegraphics[width=0.97\linewidth]{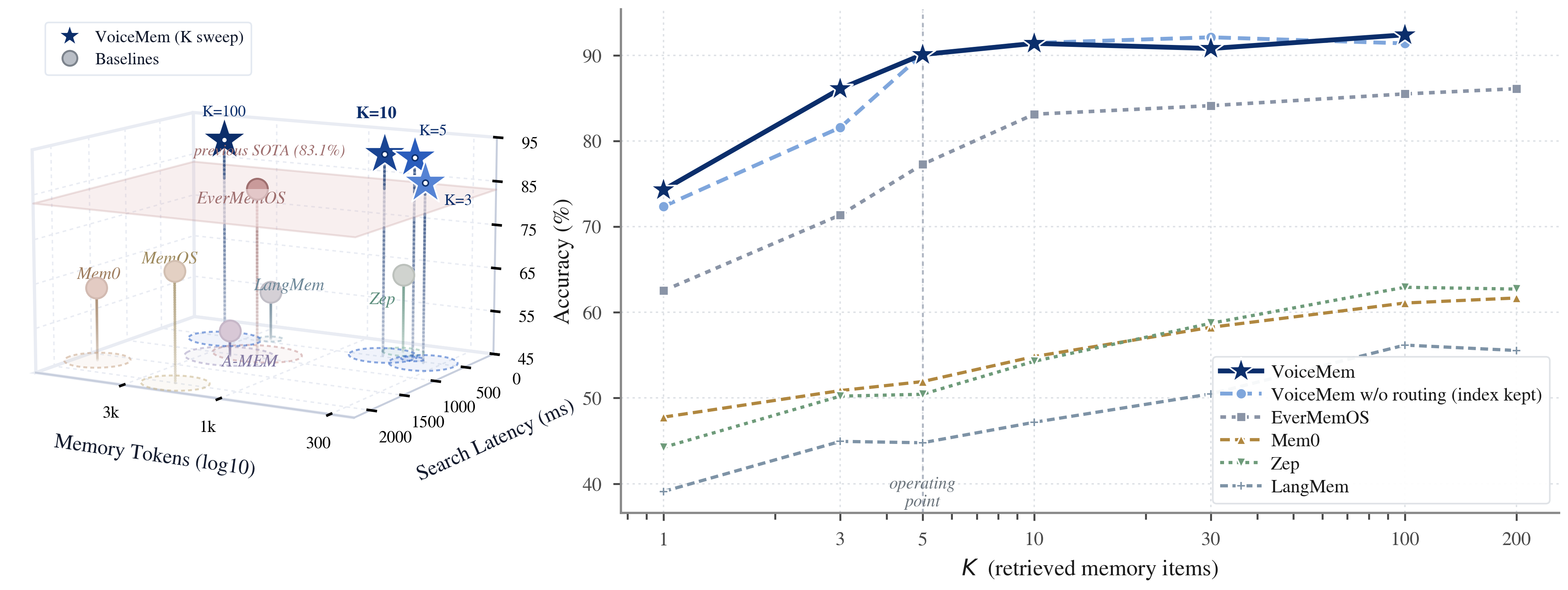}
    \vspace{-2mm}
\begin{minipage}[t]{0.42\textwidth}
    \vspace{-3mm}
    \captionof{figure}{Time--Cost--Accuracy Comparison on LoCoMo.
}
    \label{fig:frontier3d}
\end{minipage}
\hfill
\begin{minipage}[t]{0.55\textwidth}
\vspace{-3mm}
    \captionof{figure}{Retrieval-budget sweep on LoCoMo.}
    \label{fig:ksweep}
\end{minipage}
\end{figure}

\begin{figure}[t]
    \centering
    \vspace{-3mm}
\includegraphics[width=\linewidth]{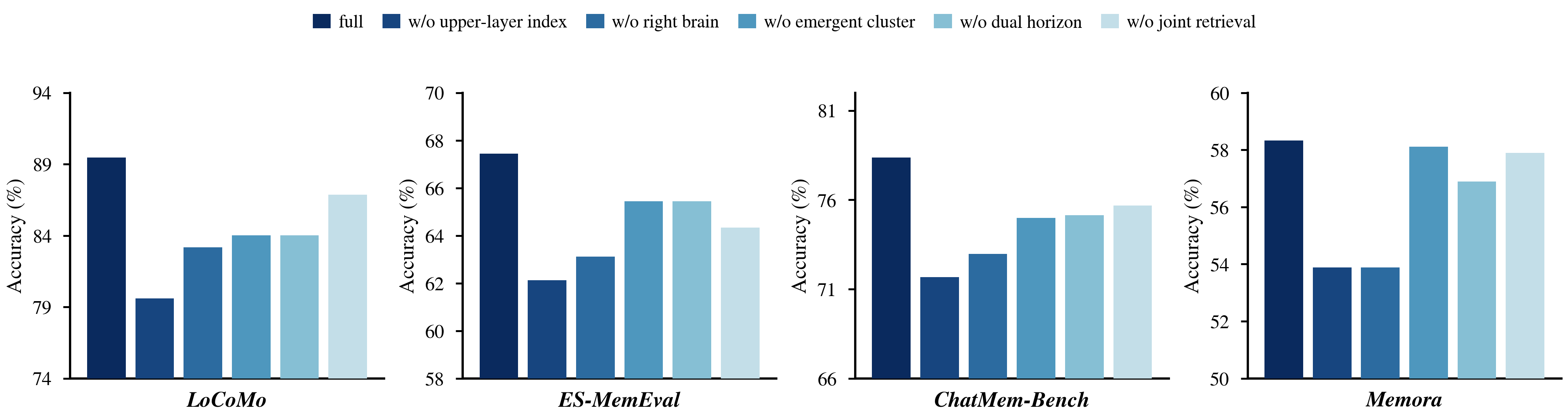}
\captionof{figure}{\textbf{Component ablation at $K{=}5$,} one mechanism disabled
per bar. Each panel uses its own $y$-range; none starts at zero.}
\label{fig:component}
\end{figure}

\noindent
\textbf{[RQ2] Accuracy, cost and latency at one operating point.}\quad
Figure~\ref{fig:frontier3d} plots all three axes at once. VoiceMem at $K{=}5$
reaches $91.2$ with $430$ memory tokens and $134$\,ms of retrieval. The strongest
baseline, EverMemOS, reaches $83.13$ with $1{,}899$ tokens: \textbf{$+8.1$ points
at $4.4\times$ fewer tokens.} The whole baseline field sits at $541$ to
$6,956$ tokens, so the saving is not specific to one comparison. Latency is
also flat in $K$: the search stays near $134$\,ms from $K{=}3$ to $K{=}100$,
\textbf{because schema routing bounds the candidate pool before ranking, so the
number of items finally returned does not change how much is searched.}

\noindent
\textbf{[RQ2] The small budget costs nothing in accuracy.}\quad
Figure~\ref{fig:ksweep} sweeps $K$ on LoCoMo. \textbf{VoiceMem is highest at
every $K$, and the margin is widest where the budget is tightest:} 
$+11.8$ over
EverMemOS and $+26.5$ over Mem0 at $K{=}1$, and $+12.8$ over EverMemOS at
$K{=}5$. The curve is flat past $K{=}5$ --- $K{=}10$ adds $1.3$ points, $K{=}100$
adds $2.3$ for $8\times$ the tokens --- so \textbf{operating at $K{=}5$ gives up
essentially nothing.} The baselines have no such option: EverMemOS needs top-$10$ to reach $83.13$,
which VoiceMem already exceeds at $K{=}3$ with $362$ tokens, and neither Mem0
nor Zep exceeds $63$ at any budget. The dashed curve removes schema routing and shows where the saving
comes from: it costs $4.53$ points at $K{=}3$ and closes to within $0.05$ by
$K{=}5$, i.e.\ \textbf{routing does not raise the ceiling, it lowers the budget
needed to reach it} --- without it, matching $430$-token accuracy takes $K{=}30$
and $1{,}277$ tokens. Note that this ablation disables schema-based routing while
\emph{retaining} the upper-layer index, so the candidate pool is still narrowed;
removing the index entirely costs $9.9$ points at the same $K{=}5$
(Figure~\ref{fig:component}), which is what actually carries the gain.

\noindent
\textbf{[RQ3] Every mechanism contributes on every dataset.}\quad
Figure~\ref{fig:component} disables one mechanism at a time on four datasets.
All five ablations lose accuracy on all four. Removing the upper-layer index is
the largest loss everywhere ($-9.9$ / $-5.3$ / $-6.7$ / $-4.4$ on LoCoMo /
ES-MemEval / ChatMem-Bench / Memora). The right brain is next ($-6.3$ / $-4.3$ / $-5.4$ /
$-4.4$), so \textbf{affect and persona carry information the factual store does
not.} Emergent clustering ($-5.5$ / $-2.0$ / $-3.4$ / $-0.2$) and dual-horizon
updating ($-5.4$ / $-2.0$ / $-3.2$ / $-1.4$) are largest on LoCoMo, whose
sessions are the longest, and smallest on Memora. Joint retrieval is the smallest
on LoCoMo and ChatMem-Bench ($-2.6$ / $-3.1$ / $-2.7$ / $-0.4$): taking the union of two
separately ranked lists is a workable fallback. Two further ablations --- how a
one-hop neighbour schema should enter the prompt, and how cluster maintenance
compares against splitting at random --- are in Appendix~\ref{app:abl}.

%% ---------- 双脑散点 ----------
\begin{figure}[!t]
\centering
\includegraphics[width=\linewidth]{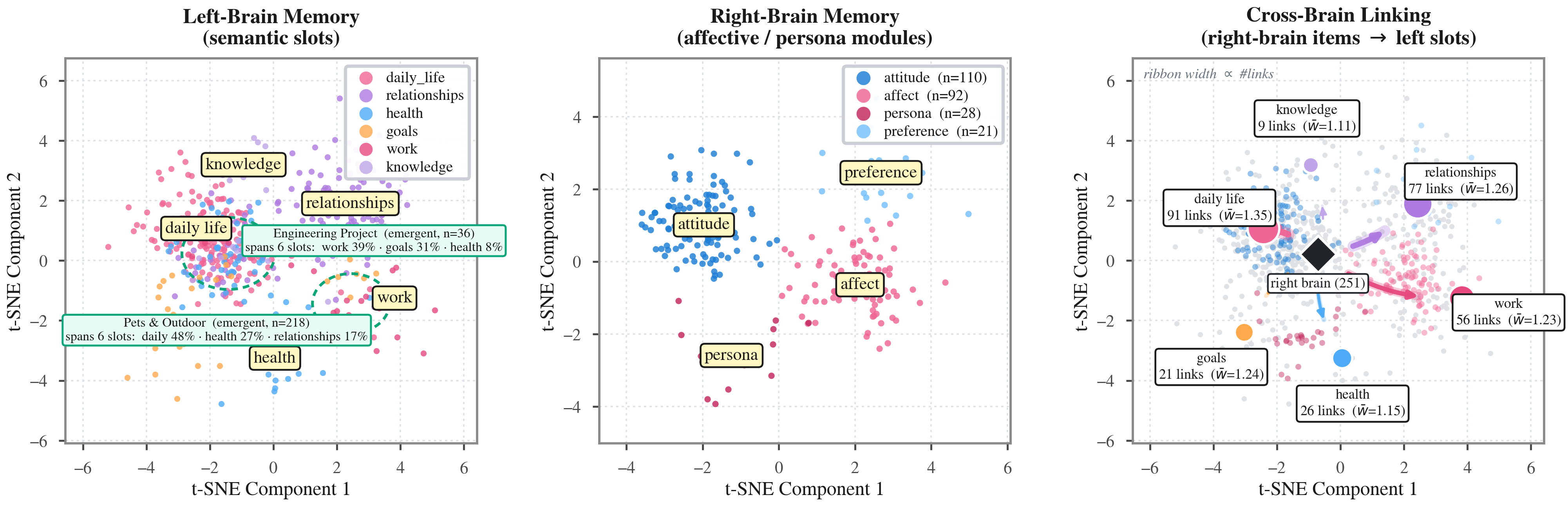}
\caption{\textbf{What the two stores hold.} \emph{Left:} left-brain items by
semantic slot, with the two emergent sub-slots circled. \emph{Middle:} right-brain
items by module. \emph{Right:} right-to-left linking, ribbon width proportional to
link count.}
\vspace{4mm}
\label{fig:dualbrain}
\end{figure}

\noindent
\textbf{[RQ3] The structure those mechanisms produce.}\quad
Figure~\ref{fig:dualbrain} shows what the store looks like once they run. The
left store holds $510$ items across six preset slots. Two further slots were
never preset; they emerged from the semantic graph and cover $254$ items, $49.8\%$
of the store. Neither sits inside a single preset slot: \emph{Pets \& Outdoor}
($218$ items) draws $48\%$ from \texttt{daily\_life}, $27\%$ from
\texttt{health} and $17\%$ from \texttt{relationships}, and \emph{Engineering
Project} ($36$ items) splits between \texttt{work} ($39\%$) and \texttt{goals}
($31\%$). Both span all six preset slots, so \textbf{emergence re-partitions the
store across preset boundaries rather than refining within them} --- which is why
disabling it costs most on LoCoMo, where sessions run longest and the preset
slots are furthest from the actual topic structure. The right store holds $251$
items in four modules, linked to left-brain slots by $280$ edges spread unevenly:
\texttt{daily\_life} receives $91$, \texttt{knowledge} $9$.

\noindent
\begin{minipage}[c]{0.58\textwidth}
\textbf{[RQ4] The index transfers across backends.}\quad
The index requires identifier-addressable items, similarity search, and \textsc{update} writes. Without threshold retuning, it improves all three stores by \textbf{$15.8$--$29.5$ points} (Table~\ref{tab:backend}), showing the gain is not store-specific. Mem0 is the full system; however, LangMem and Mem0 start within $5.50$ points but finish $19.26$ apart, so backend quality still bounds recovery.
\end{minipage}
\hfill
\begin{minipage}[c]{0.4\textwidth}
\centering
\captionof{table}{Backend transfer on LoCoMo.}
\vspace{-2mm}
\label{tab:backend}
\footnotesize
\setlength{\tabcolsep}{4pt}
\renewcommand{\arraystretch}{1.04}
\begin{tabular}{@{}l C{1.15cm} C{1.15cm} C{1.15cm}@{}}
\toprule
\textbf{Backend} & \textbf{bare} & \textbf{+ ours} & $\Delta$ \\
\midrule
Mem0 & 61.68 & \textbf{91.20} & \textbf{+29.52} \\
LangMem          & 56.18 & \textbf{71.94} & +15.76 \\
Zep              & 62.93 & \textbf{85.85} & +22.92 \\
\midrule
\textit{mean}    & 60.26 & \textit{83.00} & \textit{+22.73} \\
\bottomrule
\end{tabular}
\end{minipage}

%% 恢复本节开头临时改掉的浮动体上限
\setcounter{topnumber}{4}
\setcounter{totalnumber}{6}

\section{Case Study}
\begin{figure}[!t]
    \centering
    \includegraphics[width=1\linewidth]{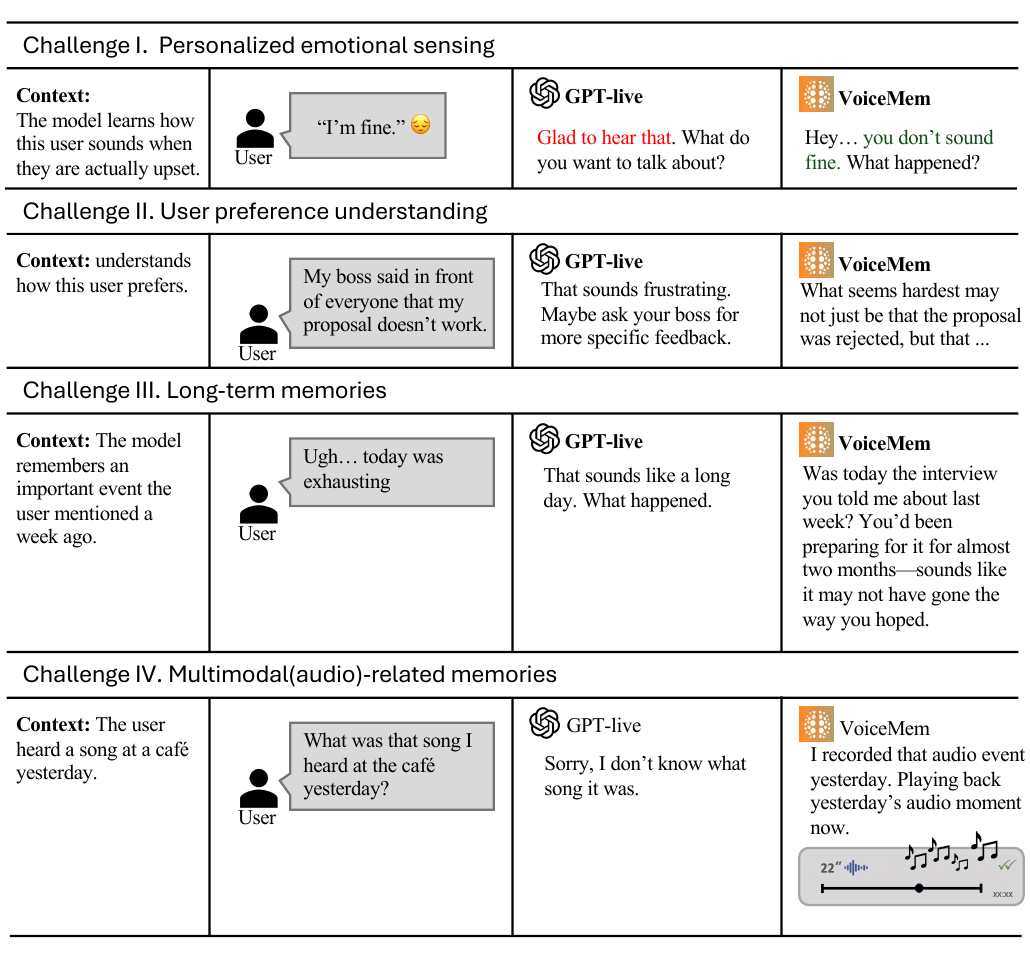}
    \caption{
    \textbf{VoiceMem augments real-time voice models with persistent, personalized memory.}
    Across four representative cases, VoiceMem enables user-specific emotion sensing, preference-aware responses, long-term recall, and multimodal audio memory that are unavailable to a memoryless baseline.
    }
    \label{fig:case_study}
\end{figure}

\vspace{2mm}
As shown in Fig.~\ref{fig:case_study}, current real-time voice models can respond fluently to the current turn but struggle to accumulate persistent knowledge about the user.
\textbf{VoiceMem} addresses this limitation by jointly modeling semantic, preference, affective, and audio-related memories, enabling responses conditioned on a user's interaction history.
Its streaming dual-brain architecture separates factual memory from affective attribution while supporting long-horizon and multimodal recall.
As a result, VoiceMem turns turn-level speech understanding into continuous, personalized user understanding with minimal latency overhead.

\vspace{1mm}
\section{Conclusion}
\vspace{2mm}
In this paper, we presented \textsc{VoiceMem}, a streaming dual-brain memory framework that
equips real-time conversational systems with both informational and emotional memory without
breaking their latency budget. The left brain organizes factual knowledge through a two-level
schema--entity index with a query-driven cluster emergence mechanism, keeping the candidate pool
dense enough to survive a \textbf{top-5} retrieval budget, while the right brain models the person through
independent and cross-entity persona nodes maintained by short- and long-horizon affective
attribution. A four-stage streaming query then hides the entire retrieval inside the silence that a
standard VAD already waits out. Around this core we built a complete pipeline for memory-aware
SLM training, long-horizon evaluation, and decoupled deployment with interchangeable backends.
Experiments show that \textsc{VoiceMem} leads on both information and persona benchmarks at a
retrieval budget an order of magnitude below common practice, reaching \textbf{91.2} on LoCoMo with
\textbf{430} memory tokens and \textbf{134\,ms} of retrieval, and that the same index lifts three
different backends by \textbf{15.8--29.5} points. By turning turn-level speech understanding into
continuous, personalized user understanding at essentially no latency cost, \textsc{VoiceMem} lays a
practical memory foundation for the next generation of empathetic, real-time voice assistants.

\newpage
\bibliographystyle{plainnat}
\bibliography{sample}

% ==================================================================
%%  附录 —— 如果 main.tex 里已经有 \appendix，把下面这一行删掉
%% ==================================================================
\newpage
\appendix

\begin{center}
{\LARGE\bfseries Appendix}
\end{center}
\vspace{2mm}

\section{Related Work}
\label{sec:related}
\paragraph{Agent memory.}
Agent memory has evolved from scalable factual stores such as Mem0~\citep{mem0} and Zep~\citep{zep}, to self-organizing and graph-structured systems that revise links or consolidate memories over time~\citep{a-mem,apexmem2026,jiang2026magma,wu2026gam}. A parallel line incorporates affect into retrieval and updating, including Emotional RAG~\citep{emotionalrag}, KEEM~\citep{keem}, dynamic affective memory management~\citep{lu2025dynamic}, and DualMem~\citep{dualmem2026}. Yet these methods remain largely text-centric: affect mostly re-weights stored content, while factual relations and person-directed attribution are not maintained as distinct evolving states. VoiceMem instead separates an accumulating relational-information memory from a revising persona memory (Sections~\ref{sec:leftbrain} and~\ref{sec:right}).

\paragraph{Speech language models.}
Speech language models have progressed from conversational overlap to full-duplex interaction and reasoning during speech. dGSLM~\citep{nguyen2022dgslm} models overlapping dialogue, while LSLM~\citep{lslm}, SyncLLM~\citep{veluri2024syncllm}, Mini-Omni~\citep{xie2024mini}, Mini-Omni2~\citep{xie2024miniomni2}, Moshi~\citep{moshi}, SALMONN-omni~\citep{salmonn2024omni} and Audio-Interaction~\citep{xie2026audio} progressively support simultaneous listening, speaking, and low-latency multimodal generation. More recent systems move reasoning into the interaction window: Mind-Paced Speaking~\citep{mps2025}, SHANKS~\citep{shanks2025}, Chronological Thinking~\citep{chronological2025}, and Mini-Omni-Reasoner~\citep{xie2025miniomnireasoner} interleave reasoning with ongoing speech. However, memory remains largely outside this real-time path; VoiceMem extends the streaming constraint to memory access and revision themselves (Section~\ref{sec:stream}).

\paragraph{Memory-interaction systems.}
A complementary line studies how internal state shapes interactive behavior. Neuroscience suggests specialized parallel pathways for speech~\citep{hickok2007dual} and distinct modulation of emotionally salient memory~\citep{mcgaugh2004amygdala}, while computational dual-system models such as Talker-Reasoner~\citep{christakopoulou2024talker} and Fast-in-Slow~\citep{fastinslow2025} separate fast interaction from deeper reasoning. Proactive agents further use inferred intent or persistent context to decide when and how to act~\citep{zhang2024proagent,lu2025proactive}; PASK~\citep{xie2026pask}, in particular, combines proactive interaction with long-term memory. These systems use memory or internal state to support reasoning and action selection, whereas VoiceMem focuses on the memory substrate itself: factual-relational and persona-affective states are maintained separately and updated in parallel during continuous interaction.

\section{Additional Ablations}
\label{app:abl}
 
\subsection{One-hop neighbours: compress or expand}
\label{app:macro}
 
When a schema has a strongly linked neighbour, that neighbour can either be
compressed into a one-sentence description or expanded into memory items that
compete for the top-$K$ slots. Table~\ref{tab:macro} compares both against
ignoring neighbours entirely.
 
Compression and macro-expansion are indistinguishable on LoCoMo ($91.2$ against
$91.10$), but compression is clearly better on ES-MemEval ($+2.76$), and ignoring
neighbours costs $3.4$ and $9.1$ respectively --- so the neighbour information is
doing real work either way. What separates the two is the candidate pool:
compression keeps it from growing ($144.1$ against $165.7$ on ES-MemEval) at the
price of $71.2$ extra prompt tokens on LoCoMo. \textbf{We take compression because
it bounds the pool, not because it is more accurate.}
 
\begin{table}[h]
\centering
\caption{\textbf{How a one-hop neighbour schema should enter the prompt.}
\emph{Pool} is the mean number of items entering final ranking.}
\label{tab:macro}
\small
\setlength{\tabcolsep}{5pt}
\renewcommand{\arraystretch}{1.08}
\begin{tabular}{@{}ll C{1.35cm} C{1.25cm} C{1.45cm}@{}}
\toprule
\textbf{Dataset} & \textbf{Neighbour handling}
  & \textbf{Acc.} & \textbf{Pool} & \textbf{Mem. tok.} \\
\midrule
\multirow{3}{*}{LoCoMo}
  & description \emph{(ours)} & \textbf{91.2} & 292.0 & 430 \\
  & macro-expansion           & 91.10 & 306.0 & \textbf{358.8} \\
  & w/o macro                 & 87.80 & \textbf{292.0} & 427.2 \\
\midrule
\multirow{3}{*}{ES-MemEval}
  & description \emph{(ours)} & \textbf{67.45} & \textbf{144.1} & 624.4 \\
  & macro-expansion           & 64.69 & 165.7 & \textbf{497.6} \\
  & w/o macro                 & 58.39 & 144.8 & 498.1 \\
\bottomrule
\end{tabular}
\end{table}
 
\subsection{Cluster maintenance over a growing history}
\label{app:emerge}
 
Table~\ref{tab:emerge} compares four cluster-maintenance strategies on
ES-MemEval P1 ($32$ sessions), scored over all $252$ questions at $K{=}5$.
 
Emergence is the most accurate at $74.40$, ahead of static by $1.80$ points, of
\texttt{random\_split} by $2.00$ and of \texttt{size\_threshold} by $2.79$.
\textbf{The comparison that carries the argument is emergence against
\texttt{random\_split}}, which is forced to perform the same number of splits:
the $2.00$-point gap means the gain comes from splitting in the right place, not
from splitting at all. Splitting in the wrong place is worse than not splitting
--- \texttt{size\_threshold} falls $0.99$ points \emph{below} static, and
\texttt{random\_split} does not beat it either.
 
\begin{table}[h]
\centering
\caption{\textbf{Cluster maintenance strategies} on ES-MemEval P1, $K{=}5$,
weighted over all $32$ sessions ($n{=}252$). \texttt{random\_split} is forced to
perform the same number of splits as \texttt{emergence}.}
\label{tab:emerge}
\small
\setlength{\tabcolsep}{6pt}
\renewcommand{\arraystretch}{1.08}
\begin{tabular}{@{}l C{2.4cm} C{2.2cm}@{}}
\toprule
\textbf{Strategy} & \textbf{Acc.} & $\Delta$ \textbf{vs.\ static} \\
\midrule
static \emph{(no split)}   & 72.60 & --- \\
size\,threshold            & 71.61 & $-0.99$ \\
random\,split              & 72.40 & $-0.20$ \\
\midrule
\textbf{emergence \emph{(ours)}} & \textbf{74.40} & \textbf{$+1.80$} \\
\bottomrule
\end{tabular}
\end{table}

\end{document}